\documentclass[twocolumn,showpacs,preprintnumbers,amsmath,amssymb,superscriptaddress]{revtex4}

\usepackage[maxfloats=256]{morefloats}
\usepackage{graphicx}
\usepackage{dcolumn}
\usepackage{bm}
\usepackage{tabularx}
\usepackage{ulem}
\newcolumntype{M}{>{\centering\arraybackslash}m{1.85cm}}
\usepackage[export]{adjustbox}
\usepackage{float}
\usepackage{color}   
\usepackage{hyperref}
\hypersetup{
	colorlinks=true, 
	linktoc=all,     
	linkcolor=blue,  
}

\makeatletter
\newcommand{\colorcaption}[2][]{%
	\begingroup%
	\renewcommand{\@caption@fignum@sep}{ (Color online). }%
	\caption[#1]{#2}%
	\endgroup%
}
\makeatother

\begin{document}
\title{ Nuclear structure properties of $^{193-200}\mathrm{Hg}$ isotopes within large-scale shell model calculations}
	\author{ Subhrajit Sahoo}
	\address{Department of Physics, Indian Institute of Technology Roorkee, Roorkee 247667, India}
	\author{ Praveen C. Srivastava\footnote{Corresponding author: praveen.srivastava@ph.iitr.ac.in}}
	\address{Department of Physics, Indian Institute of Technology Roorkee, Roorkee 247667, India}
	\author{Noritaka Shimizu}
	\address{Center for Computational Sciences, University of Tsukuba, 1-1-1, Tennodai Tsukuba,\\ Ibaraki 305-8577, Japan}
	\author{Yutaka Utsuno}
	\address{Advanced Science Research Center, Japan Atomic Energy Agency, Tokai, Ibaraki 319-1195, Japan}
    \address{Center for Nuclear Study, University of Tokyo, Hongo, Bunkyo-ku, Tokyo 113-0033, Japan}

\begin{abstract}
Large-scale shell-model calculations have been performed to study the nuclear structure properties of Hg isotopes with mass varying from $A=193$ to $A=200$. The shell-model calculations are carried out in the 50 $\leq Z \leq$ 82 and 82 $ \leq N \leq$ 126 model space using monopole-based truncation. We present detailed studies on low-energy excitation spectra, energy systematics, and collective properties of Hg isotopes, such as reduced transition probabilities, quadrupole, and magnetic moments along the isotopic chain. The evolution of wave function configurations with spin is analyzed in the case of even-$A$ Hg isotopes.
The shell-model results are in reasonable agreement with the experimental data and predictions are made where experimental data are unavailable. The shapes of Hg isotopes are also investigated through the energy-surface plots.
\end{abstract}


\maketitle

\section{Introduction}
 The nuclei lying in the neutron-deficient Pb region ($Z \leq 82$ and $N \leq 126$) display various structural phenomena. The Hg isotopes are some of the clearest examples to understand these different structural properties. 
The optical spectroscopy experiments have identified a large difference in the isotopic shifts between odd and even Hg isotopes known as shape staggering around the mass region $A \sim 181$--$188$ \cite{shape_stagering, ch_radiiHg, shape_stageringNat}. The shape coexistence features are are also observed in the neutron-deficient Hg isotopes \cite{KHyde_rev,sh_coExp,sh_coIBMCM} with a minimal effect in $^{190}$Hg \cite{Hgdata2019}. The structural behaviors of Hg isotopes with mass $ A \geq 192$ are crucial in both experiments and theory. Unlike the lighter Hg isotopes, the mixing between normal and intruder configurations is significantly reduced in these isotopes. They exhibit weakly deformed ground-state configuration, and a spherical shape is attained as we move towards the $N=126$ shell closure. Extensive studies have been done on the structural properties of these isotopes in recent years. The lifetimes of yrast $2_1^+$, $4_1^+$ and some negative parity states were determined in the even-even $^{192-196}$Hg isotopes using $\gamma$-$\gamma$ fast timing spectroscopy and reduced transition probabilities ($B(E2)$ values) were derived for transition between these states \cite{Hgdata2018}. This work was further extended up to $^{200}$Hg by Olaizola and his collaborators, which covers $B(E2)$ measurements of yrast and some non-yrast states \cite{Hgdata2019}. Further, the collective properties and half-lives of the isomeric states were measured in the even-even Hg isotopes up to high spin states by Suman \textit{et al.} in \cite{Hgdata2021}. The experimental studies on odd mass Hg isotopes are available in Refs. \cite{Hg193,Hg195,Hg197_199,Proetel-odd-Hg-1974}.

The nuclear shell model (SM) has been incredibly successful in explaining various structural properties of nuclei in different mass regions, such as binding energies, excitation spectra, band structures, isomers, electromagnetic transition rates, $\beta$ decay properties, and many more \cite{SM_RevBrown, SM_RevCaurier, Bbhoy1, Bbhoy2, Bbhoy3,yoshinaga201_206, suzuki206_208, Akumar1, Akumar2, betadecayYoshida,yanase_Shiff_2020}. 
However, the heavy mass systems (particularly the mid-shell nuclei $Z \approx 82$ and $N \approx 104$) in the full configuration space are beyond the reach of the state-of-the-art shell-model calculation. Hence, other models like the Interacting Boson Model (IBM) and the Density Functional Theory (DFT) have been mainly used to study the nuclear properties of such systems. Some recent articles where the low energy excitation spectra, transition strengths, and deformation properties of the whole Hg chain were addressed in the framework of IBM can be found in \cite{sh_coIBMCM, IBM2}. The electromagnetic moments of Hg nuclei were calculated using DFT in \cite{DFT_Hg}.
With the advancement of computational facilities, it is also possible to perform a shell-model study of the Hg isotopes in the large model space by applying suitable truncation. Recently, the shell-model calculations have been performed for Hg isotopes lying towards $N=126$ shell closure ($^{206-208}$Hg), and the energy spectra, $E2$ strengths, quadrupole, and magnetic moments are compared with experimental data \cite{suzuki206_208}. Similar investigations have been done to study the nuclear properties of $^{201-206}$Hg in \cite{yoshinaga201_206}. As we go further below the $N=126$ shell closure, the dimension of the shell-model Hamiltonian matrix is too huge to be diagonalized, and thus the truncation of the model space is unavoidable.
Hence, in this article, we aim to study the nuclear structure properties of Hg isotopes below $A \leq 200$. Motivated by the availability of numerous experimental data for the Hg isotopes in the region $190 \leq A \leq 200$, the objective of this work is to provide a shell-model estimate of these measured observables.  In addition, a shell-model calculation of $^{199}$Hg was achieved recently to evaluate the nuclear Schiff moment, which is a key quantity for experiments aimed at detecting the permanent atomic electric dipole moment ~\cite{yanase_Shiff_2020}. It is, therefore, crucial to test the validity of the shell model and its effective interaction in this mass region systematically.  
To the best of our knowledge, we are reporting the systematic large-scale shell-model calculations of $^{193-200}$Hg for the first time.

The present work is organized as follows: Sec.~\ref{sec2} contains information regarding the model space and interaction used in the shell-model calculations. The shell-model results of even-$A$ Hg isotopes followed by odd-$A$ Hg isotopes are presented in Sec.~\ref{sec3}. The results include a detailed analysis of low-energy excitation spectra, electromagnetic transition rates, quadrupole, and magnetic moments of the Hg isotopes. 
The energy surfaces of Hg isotopes given by the Q-constrained Hartree-Fock-Bogoliubov calculations are also discussed. 
The conclusions are summarized in Sec.~\ref{sec5}.

\section{Theoretical framework}
\label{sec2}

\subsection{Model space and interaction}

To study the Hg isotopes, the shell-model calculations were carried out using the KHHE \cite{Khhe} interaction. The matrix elements of the KHHE interaction are based on the realistic effective Kuo-Herling (KHH) interaction \cite{Kuo1,Kuo2}. The Kuo-Herling interaction was derived from the realistic Hamada-Johnston potential \cite{Hamada} for particles below and above $N = 126$ shell closure. The KHHE interaction was constructed for the model space 50 $\leq Z \leq$ 82 and 82 $ \leq N \leq$ 126 above the inert $^{132}_{50}$Sn core by implementing the modifications in proton-proton and proton-neutron interactions of KHH as suggested in the Ref. \cite{Kuomod} in order to describe the nuclei around $^{208}$Pb region more precisely. The KHHE interaction also includes further modification in the neutron-neutron interaction part \cite{Khhe}. The model space of this interaction consists of five proton orbitals ($2s_{1/2},0g_{7/2},0h_{11/2},1d_{5/2},1d_{3/2}$) and six neutron orbitals ($0h_{9/2},0i_{13/2},1f_{7/2},2p_{3/2},1f_{5/2},2p_{1/2}$).

The shell-model calculation for these Hg nuclei in the entire model space is beyond the capability of the conventional shell model diagonalization method even utilizing the latest computational facilities. To avoid this difficulty, we introduce monopole-based truncation described in the next subsection.

\subsection{Monopole-based truncation}
In the $M$-scheme shell-model calculations, the nuclear wave function is expressed as a linear combination of the vast number of Slater determinants, called the $M$-scheme basis states \cite{Kshell}.  
The number of all basis states allowed in the model space, namely the $M$-scheme dimension, is often too huge to be tractable even utilizing a state-of-the-art supercomputer. 
In that case, it is necessary to truncate the model space so that the computation is feasible.
The most widely used way of truncation is to choose the basis states whose number of particle-hole excitations across a certain shell gap is restricted. 
However, in the present case, there is no suitable shell gap for truncation especially in the neutron valence orbits.
To efficiently truncate the model space based on importance, we adopt the monopole-based truncation \cite{Qi2016,Yuan2023}. 
In this truncation scheme, we divide the $M$-scheme Slater determinants into their group which has the same occupancies $\{n_\alpha\}$ of single-particle orbitals.
This group is called ``partition'' and is a minimal set respecting rotational and parity symmetries \cite{zelevinsky1996}.  
We pick up the partitions whose energies are lower than a certain criterion so that the tractable subspace is spanned.
The energy of each portion is defined employing the monopole Hamiltonian as follows.

The two-body monopole Hamiltonian is defined as
\begin{equation}
\hat{H}_\textrm{mono}= \sum_\alpha e_\alpha \hat{n}_\alpha +  \frac12 \sum_{\alpha,\beta} 
\bar{v}_{\alpha,\beta} : \hat{n}_\alpha \hat{n}_\beta:
\end{equation}
where $\alpha$ and $\beta$ denote single-particle orbits. $e_\alpha$ and $\hat{n}_\alpha$ are the single-particle energy of the shell-model interaction and the number operator of the orbit $\alpha$, respectively \cite{otsuka2020evolution}.
``::'' denotes the normal ordering.
The matrix element of the two-body monopole interaction $\bar{v}(\alpha,\beta)$ is given by averaging the angular-momentum dependence as
\begin{equation}
	\bar{v}_{\alpha,\beta}
 =\frac{\sum_{J}(2J+1)
 \langle \alpha\beta|V|\alpha\beta\rangle_J}{\sum_{J}(2J+1)},
\end{equation}
where $\langle \alpha\beta|V|\alpha\beta\rangle_J$ is a so-called two-body matrix element (TBME) of the shell-model interaction.
Any $M$-scheme basis states belonging to a partition has the same energy expectation value of this monopole Hamiltonian.  
In the present scheme, we prepare all the possible partitions $\{n_\alpha\}$ and their monopole energies and select only the partitions whose energies are lower than a specified cutoff energy to span the truncated model space.

\begin{figure}[htbp]
\centering
    \includegraphics[width=8cm]{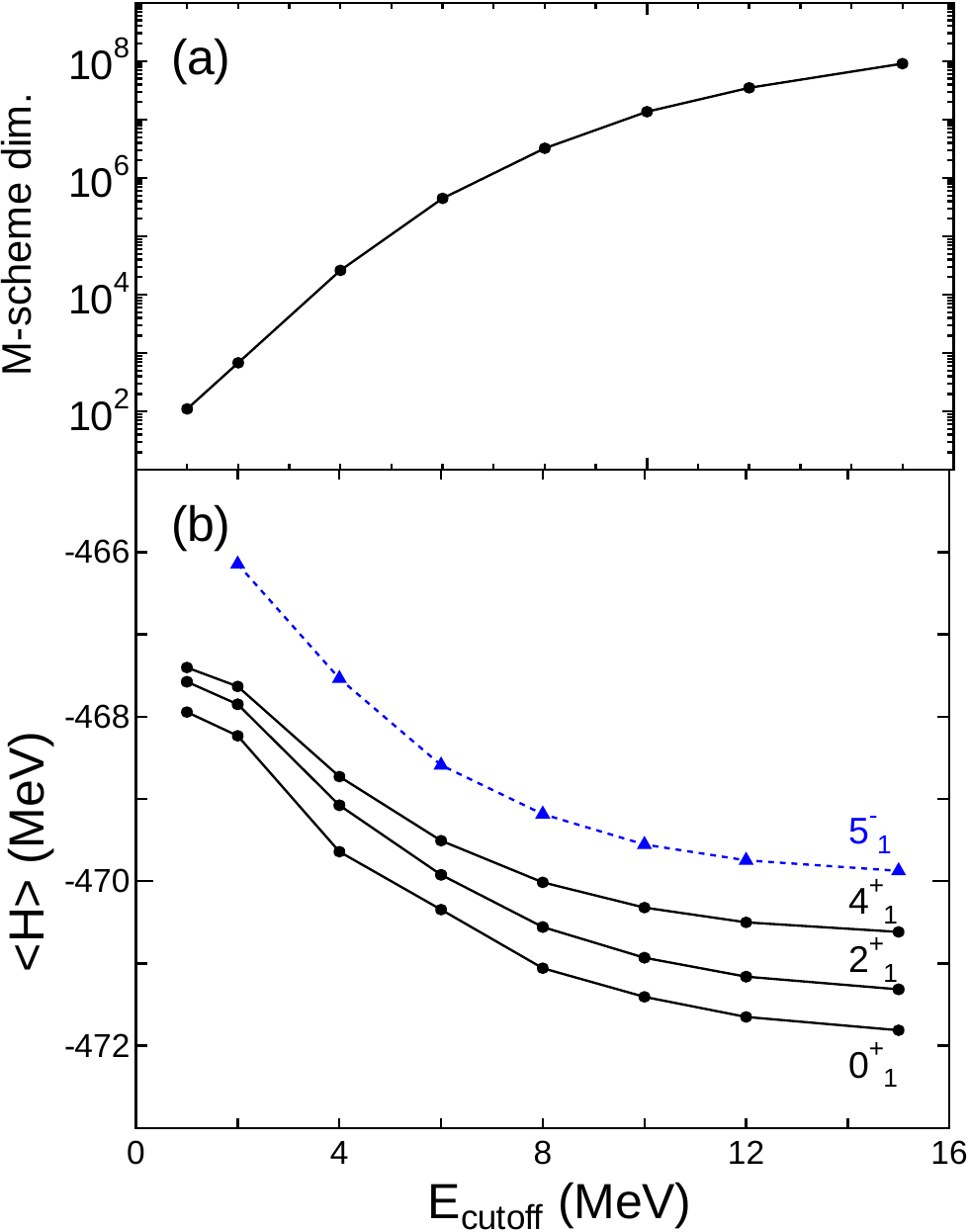}
	\caption{ (a) $M$-scheme dimension of $^{200}$Hg
 against the cutoff energy of the monopole-based truncation.
 (b) Shell-model energies of the $0^+_1$, $2^+_1$, $4^+_1$ states (black circles with solid lines) and $5^-_1$ state (blue circles with a dotted line) with the monopole-based truncation. The dimension without truncation and the corresponding exact shell-model energies are shown at $E_\textrm{cutoff}=15$ MeV.  }
 \label{fig:dim_econv}
\end{figure}

\begin{figure} 
\centering
    \includegraphics[width=8cm]{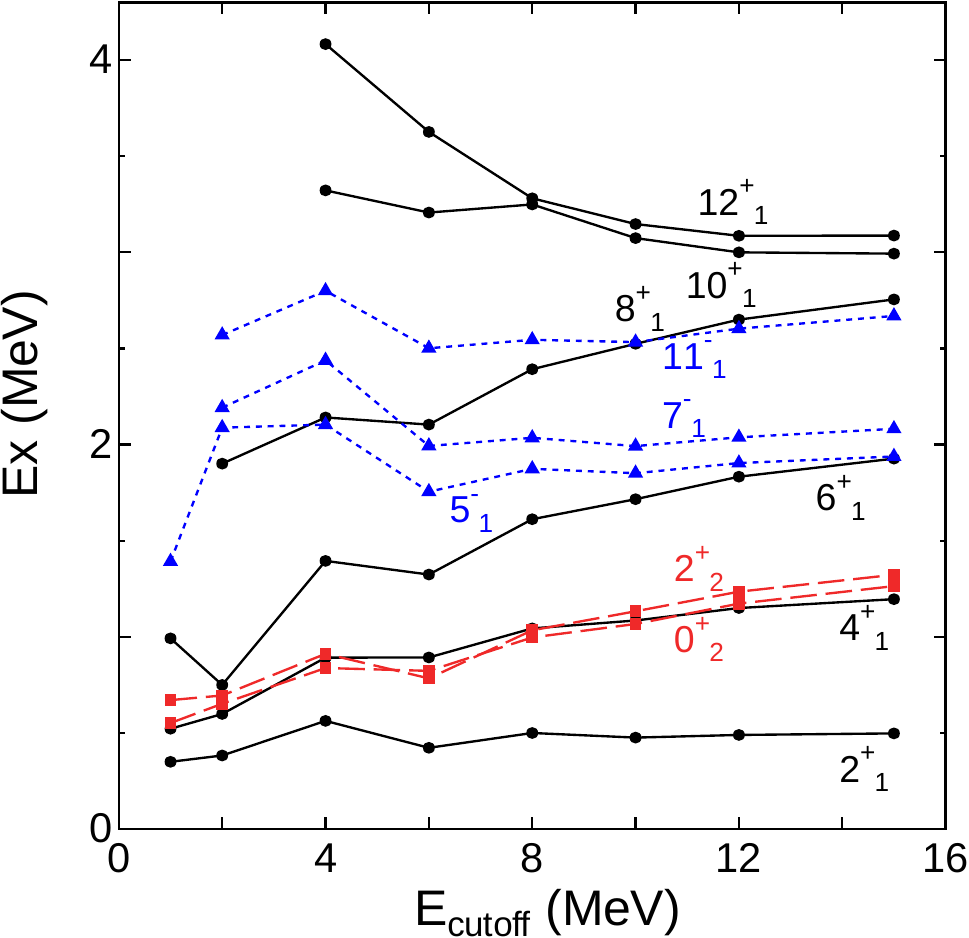}
	\caption{ Shell-model excitation energies of $^{200}$Hg with the monopole-based truncation against the cutoff energies $E_\textrm{cutoff}=1, 2, 4, 6, 8, 10, 12$ MeV.
 The black circles with the solid lines show the excitation energies of the yrast $2^+_1, 4^+_1, 6^+_1, 8^+_1, 10^+_1,$ and $12^+_1$ states. The blue triangles with the dotted lines denote those of $5^-_1, 7^-_1,$ and $11^-_1$ states.
 The squares with the dashed lines denote those of $0^+_2$ and $2^+_2$ states.
 The exact shell-model results are shown at $E_\textrm{cutoff}=15$ MeV. }
 \label{fig:ex_mbt}
\end{figure}

As a benchmark test of the monopole-based truncation, we present the shell-model results of $^{200}$Hg, the $M$-scheme dimension of which is $9.1\times10^7$, and its exact value without truncation is available. 
The shell-model code ``KSHELL" \cite{Kshell} was used for numerical computations throughout this work.
Figure \ref{fig:dim_econv} (a) shows the $M$-scheme dimension of $^{200}$Hg as a function of the cutoff energy $E_\textrm{cutoff}$, which is defined as a relative energy from the lowest one of all the partition energies. 
As the cutoff energy increases the dimension grows rapidly, and the computational cost of the shell-model calculation is almost proportional to the dimension \cite{SM_RevCaurier}.
Figure \ref{fig:dim_econv} (b) shows the convergence of the eigenenergies of the truncated model space.
The energies converge gradually as a function of the cutoff energy, while they show fast convergence as a function of the $M$-scheme dimension \cite{Qi2016}.
The excitation energies, which are the differences between these absolute energies, converge much faster than the absolute energies themselves, which will be discussed in Fig. \ref{fig:ex_mbt}. 
As a comparison with the conventional particle-hole truncation, we performed the shell-model calculation disallowing the excitation from the fully occupied $\nu0h_{9/2}$ orbit. 
Its ground-state energy is -471.081 MeV with the $M$-scheme $1.8\times 10^7$ dimension. The monopole-based truncation with $E_\textrm{cutoff}=10$ MeV provides us with the -471.406 MeV with the $1.4\times 10^7$ $M$-scheme dimension, which means that the present scheme gives the lower energy with the smaller dimension, namely a better result than the conventional truncation.
Note that the energy seems to converge exponentially as a function of $E_\textrm{cutoff}$ \cite{SM_RevCaurier,Caurier99NPA}, which may be useful for the extrapolation.

Figure \ref{fig:ex_mbt} shows the convergence of the excitation energies of $^{200}$Hg given by the monopole-based truncation. 
While these values fluctuate in the small cutoff-energy region, the reasonable convergence is shown at $E_\textrm{cutoff}=10$ MeV with the $1.3\times10^7$ dimension, which is almost 1/7 of the full space. 
In the practical calculations, we apply no truncation for $^{200}$Hg, the monopole-based truncation with $E_\textrm{cutoff}=12$ MeV for $^{198,199}$Hg, and $E_\textrm{cutoff}=10$ MeV for the other Hg isotopes.

\section{Results and discussion}
\label{sec3} 

\subsection{\textbf{Even-mass Hg isotopes}}
\label{subsec3_1}
The shell-model results for even-even Hg isotopes are presented in this section. A one-to-one comparison of the energy levels of the yrast states with the corresponding experimental data up to 3-MeV excitation energies is performed in Sec. \ref{subsec3_1_1}.
The reduced $E2$ transition strengths, electric quadrupole, and magnetic moments of the even Hg isotopes are discussed in Subsect.~\ref{subsec3_1_2} while the evolution of configurations of the positive parity yrast states are discussed Subsect.~\ref{subsec3_1_3}.

\begin{figure} 
\centering
    \includegraphics[width=90mm,height=100mm]{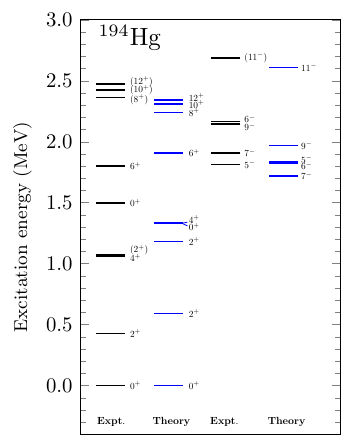}
	\caption{ Comparison between calculated and experimental \cite{NNDC} energy levels for $^{194}$Hg.}
 \label{fig:194Hg}
\end{figure}
\subsubsection{\textbf{Energy spectra}\\}
\label{subsec3_1_1}
{\bf{$^{194}$Hg:}} The calculated low-lying energy levels of $^{194}$Hg are compared with the experimental data in Fig. \ref{fig:194Hg}. The $2_1^+$, $4_1^+$, and $6_1^+$ states are observed at higher energies while the other positive parity high spin states are situated at comparatively lower energies in the calculated results. The measured spin parities of several high spin states have not yet been confirmed. In such cases, our calculated results agree with the experimental predictions. The order of these high spin states is well reproduced by the shell-model calculations. The observed energy gaps between the (8$^+$), (10$^+$), and (12$^+$) states are 0.059 and 0.052 MeV, which are close to the theoretical ones, 0.066 and 0.038 MeV. In contrast to the experimental data, the $4_1^+$ and $0_2^+$ states are degenerate.
The observed $5^-$ and $7^-$ states are also inverted in the calculated results. 

\begin{figure}[h] 
\centering
    \includegraphics[width=90mm,height=100mm]{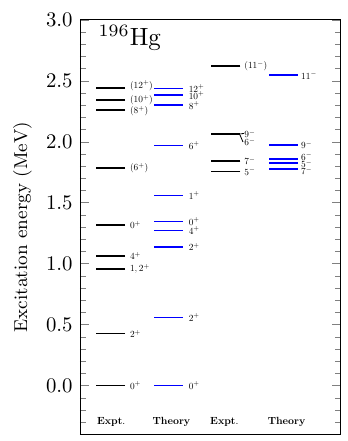}
	\caption{ Comparison between calculated and experimental \cite{NNDC} energy levels for $^{196}$Hg.}
 \label{fig:196Hg}
\end{figure}

{\bf{$^{196}$Hg:}} The experimental energy levels of $^{196}$Hg are compared with the corresponding shell-model results in Fig. \ref{fig:196Hg}.  
Similar to $^{194}$Hg, the yrast $2_1^+$, $4_1^+$, and $6_1^+$ states appeared at higher energies than the measured values. However, the energies of high spin positive parity states exhibit excellent agreement with the experimental data and are found within an energy difference of less than 40 keV. The experiment predicts a state with tentative spin 1 and 2$^+$ at the same energy 0.958 MeV, but in the calculated energy levels, the 1$^+_1$ and 2$^+_2$ states appear at energies above 1 MeV. An inversion is observed between the $5^-$ and $7^-$ states as found in $^{194}$Hg. But the energy gaps between the yrast $5^-$, $7^-$, $9^-$, and $11^-$ states are in good agreement with the experimental data.

\begin{figure}[h] 
\centering
    \includegraphics[width=90mm,height=100mm]{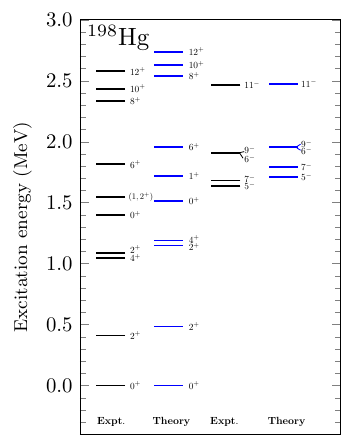}
	\caption{ Comparison between calculated and experimental \cite{NNDC} energy levels for $^{198}$Hg.}
 \label{fig:198Hg}
\end{figure}
{\bf{$^{198}$Hg:}}
The low-energy experimental and calculated positive and negative parity states of $^{198}$Hg are shown in Fig. \ref{fig:198Hg}. In the shell-model results, all the yrast states appear in the same order as found in the experiment. The $8_1^+$, $10_1^+$, and $12_1^+$ states appear at higher energies but the energy differences among them are approximately similar to those observed in the experimental data. The negative parity states also show fair agreement with the measured data. Unlike $^{194,196}$Hg, the 6$^-$ and 9$^-$ states are degenerate, consistent with the findings in the experiment.

\begin{figure}[h] 
\centering
    \includegraphics[width=90mm,height=100mm]{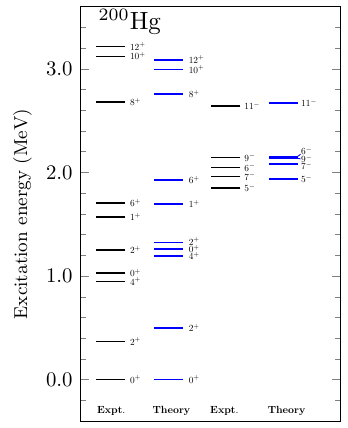}
	\caption{ Comparison between calculated and experimental \cite{NNDC} energy levels for $^{200}$Hg.}
 \label{fig:200Hg}
\end{figure}
{\bf{$^{200}$Hg:}} Figure \ref{fig:200Hg} shows a comparison of the observed low-lying energy spectra of $^{200}$Hg and the corresponding shell-model calculated results. 
The $4_1^+$, $0_2^+$, and $2_2^+$ states lie adjacent to each other suggesting surface vibrational modes.
The negative parity states also exhibit reasonable agreement with the experimental data. In contrast to the experiment, the energy gap between the $7_1^-$ and $9_1^-$ states is reduced, and the $6_1^-$ state also appears almost degenerate with the $9_1^-$ state in the calculated spectra.

\begin{figure*} 
    \includegraphics[width=80mm,height=55mm]{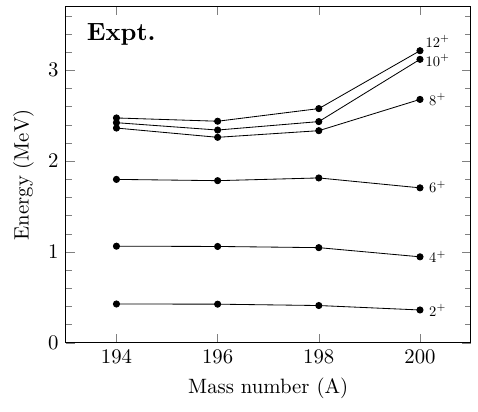}
    \hspace{-0.2cm}
    \includegraphics[width=80mm,height=55mm]{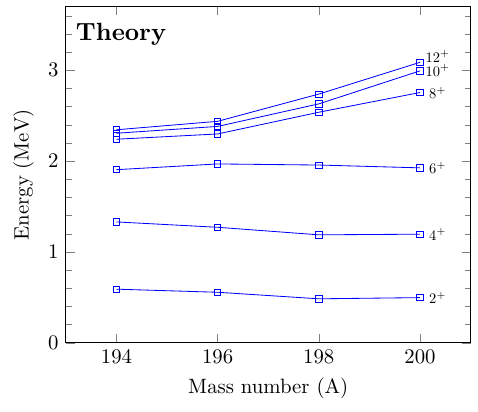}

\caption{ Experimental (left) and calculated (right) energy systematics of the positive parity yrast states in even-$A$ Hg isotopes.}
\label{fig:EvenA_posSym}
\end{figure*}

A comparison between the experimental and the shell-model results for the excitation energy systematics of positive parity yrast states is shown in Fig. \ref{fig:EvenA_posSym}. The calculated excitation energy systematics show the same behavior as the observed ones. The positive parity states show regular behavior like rotational band structure up to  6$^+$ state while the energy differences between 8$^+$, 10$^+$, and 12$^+$ states are compressed both in experimental and calculated energy spectra. A sudden change in the regular behavior of the excitation energies of the positive parity yrast states can be seen at $^{200}$Hg. The 8$^+_1$, 10$^+_1$, and 12$^+_1$ states appear at higher energies in $^{200}$Hg compared to other Hg isotopes. This is because of the $\nu 0i_{13/2}$ subshell closure observed in $^{200}$Hg \cite{200Hg}. In the Nilsson energy-level picture, a large energy gap appears in $^{200}$Hg implying $\nu 0i_{13/2}$ sub-shell closure \cite{200Hg} over which the ground-state band is built while the $10_1^+$ and $12_1^+$ states appear due to two quasi-neutron excitations of $\nu 0i_{13/2}$ sub-shell. Unlike other Hg isotopes, in $^{200}$Hg the 8$^+$ state is established as a member of the ground state band \cite{200Hg,198_200Hg} though it appears at higher excitation energy. In the dominant shell-model wave function configurations of 8$^+$ state, the $\nu 0i_{13/2}$ orbital is completely occupied. Further, a weak $E2$ transition probability from 10$^+$ to 8$^+$ state (discussed in the next Sec. \ref{subsec3_1_2}) supports the interpretation of 8$^+$ state as the ground-state band member.

\begin{figure*} 
    \includegraphics[width=80mm,height=55mm]{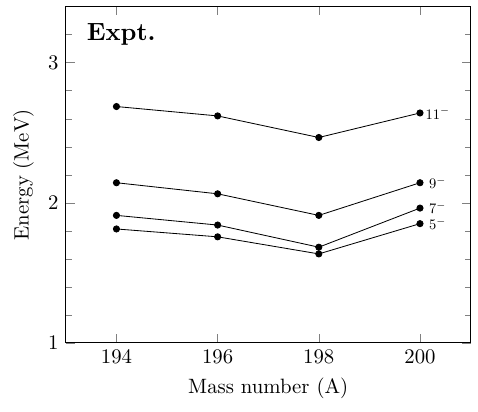}
    \hspace{-0.2cm}
    \includegraphics[width=80mm,height=55mm]{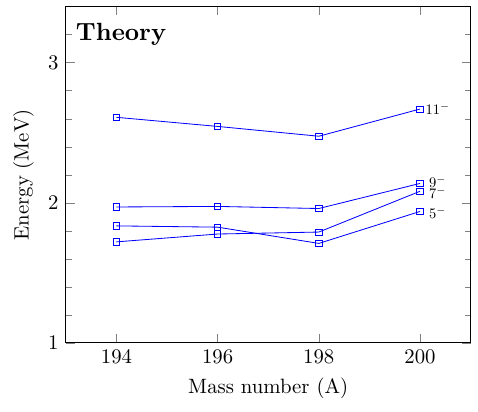}

\caption{ Experimental (left) and calculated (right) energy systematics of the negative parity yrast states in even-$A$ Hg isotopes.}
\label{fig:EvenA_negSym}
\end{figure*}

A similar comparative analysis is presented for the experimental and shell-model results on the excitation energy systematics of the negative parity yrast states in Fig. \ref{fig:EvenA_negSym}. The crossing of the lines connecting the $5^-$ and $7^-$ states indicates that these states are inverted in $^{194,196}$Hg (see Fig. \ref{fig:194Hg} and \ref{fig:196Hg}). The different cutoff energies used for monopole-based truncation in different Hg isotopes could not be the cause for such inversion. Since at lower $E_\textrm{cutoff}$, these two states do not invert in $^{200}$Hg (Fig. \ref{fig:ex_mbt}). The shell-model calculations of $^{198}$Hg with $E_\textrm{cutoff}$= 10 MeV also reproduce the correct order of the $5^-$ and $7^-$ states. Hence, it may be within the context of the present interaction used here that the correct order of the $5^-$ and $7^-$ states are not reproduced in $^{194,196}$Hg. The experimental excitation energies of the negative parity yrast states decrease with an increase in $A$ along the chain and experience a sudden rise at $^{200}$Hg. In the shell-model results, a slight change is observed in the excitation energies of these states along the even-$A$ Hg chain, but as found in the experiment, these states appeared at higher excitation energies in $^{200}$Hg compared to other isotopes.

The $R_{4/2}=E(4^+)/E(2^+)$ ratio is an important observable used for studying the collective properties of nuclei. It provides information about the nature of the yrast $2^+$ and $4^+$ states. In every nucleus, the yrast $2^+$ state is always a ground-state band member, while the $4^+$ state may not necessarily be a member of the ground-state band. This behavior is seen in nuclei exhibiting shape coexistence (e.g., $^{182-186}$Hg) where the ratio $R_{4/2}$ is less than 2. In the present case, the experimental value of $R_{4/2}$ in the even-even nuclei fluctuates between 2.45 to 2.55  indicating the $\gamma$-soft rotor nature of the nuclei \cite{Hgdata2018}. The $R_{4/2}$ values by the shell model lie within $\sim$ 2.25 - 2.45. 
Extensive theoretical and experimental studies have been done on the nuclear properties of the lighter Hg isotopes exhibiting shape coexistence \cite{sh_coIBMCM,IBM2}. The energy systematic of the heavier transitional Hg isotopes (190 $\le  A \le$ 200) differ from the lighter ones \cite{200Hg}. The complete comparison of the measured nuclear properties of these transitional isotopes in the shell-model framework requires the large-scale computations of $^{190,192}$Hg, which are excluded here because of computational complexities. This aspect should be considered by applying further approximations or truncation and is left for future studies.
  
\subsubsection{\textbf{Electromagnetic properties}\\}
\label{subsec3_1_2}

The shell-model results for reduced electric transition strengths or $E2$ strengths, quadrupole, and magnetic moments of various states in the even-A Hg isotopes are discussed in this section. In the shell model, depending upon the size of the core, i.e., for a heavier core, larger effective charges for proton and neutron ($e_p$ and $e_n$) are preferred due to the stronger core polarization effect. A systematic investigation has been performed on the values of effective charges in the Ref. \cite{suzuki206_208} for the nuclei lying around the $N = 126$ shell gap. Quadrupole moments and $E2$ strengths of several states of Pb and Hg nuclei are considered for the investigation, and two sets of effective charges have been proposed. These two sets of $e_p$ and $e_n$ well reproduce the observed electromagnetic properties with minor differences in their result. Similar investigations have been done for effective g factors for the nuclei of this region \cite{suzuki206_208}, and six sets of g factors are reported by fitting a large number of the magnetic moments data.   
In the present study we adopt $e_p = 1.8e$ and $e_n=0.8e$ and the effective g factors for proton and neutron as $g_l^p = 1.026, g_s^p = 3.757$ and $g_l^n = 0.0 , g_s^n = -2.005$ respectively following Ref. \cite{suzuki206_208}. These values reproduce the experimental results quite well in the case of Hg isotopes.

The calculated $E2$ strengths are compared with the experimental data in Table \ref{t_be2Even}. 
First, we examine the results for $^{194,196}$Hg. Their $B(E2; 2^+_1 \to 0^+_1)$ and $B(E2; 4^+_1 \to 2^+_1)$ values are in reasonable agreement with the data. 
The shell model predicts weak $E2$ strengths for the transitions $6^+_1 \rightarrow 4^+_1$ and $(8^+_1) \rightarrow 6^+_1$ in $^{194}$Hg,  
whereas the corresponding values for the $6^+_2$ state are moderate. This indicates that the calculated $6^+_1$ state is an intruder state, i.e., a state dominated by different configurations from the ground band. 
Such an intruder state is calculated to appear as the $6^+_2$ state in $^{196}$Hg. 
It is interesting to measure the $B(E2; 6^+_1 \rightarrow 4^+_1)$ and $B(E2; (8)^+_1 \rightarrow 6^+_1)$ values in $^{194,196}$Hg 
to probe the predicted structure change.
The $B(E2)$ values for the transitions $(10)^+_1 \rightarrow (8)^+_1$ and $(12)^+_1 \rightarrow (10)^+_1$ are well reproduced for $^{194,196}$Hg.
The $B(E2)$ values are small for the transitions between the states $9^-_1$ $\rightarrow$ $7^-_1$ $\rightarrow$ $5^-_1$ compared to the experimental data.

\begin{table*}
\centering
\caption{ Shell-model $E2$ strengths (in W.u.) of the even-even Hg isotopes in comparison with the experimental values \cite{Hgdata2019, Hgdata2018, NNDC}. The results are also compared with those obtained from IBM-CM \cite{sh_coIBMCM}. }
\label{t_be2Even}
\begin{tabular}{ lccccccc } 
 \hline
 \hline
 &  \hspace{7cm} & \multicolumn{3}{c}{$B(E2)$}       \\
 \cline{3-5}
Nucleus &  \hspace{0.5cm} $J^{\pi}_i \rightarrow J^{\pi}_f $&  \hspace{0.3cm} Expt. & \hspace{0.3cm} SM  & \hspace{0.3cm} IBM-CM \cite{sh_coIBMCM} \\
 \hline 
 \hline

\\          

$^{194}$Hg   & $2^+_1 \rightarrow 0^+_1$       &  30(2)          & 25.5 & 26    \\
             & $4^+_1 \rightarrow 2^+_1$       &  $>$27          & 16.0 & 36\\
             & $6^+_1 \rightarrow 4^+_1$       &  NA             & 2.1  & 41\\
             & $(8^+_1) \rightarrow 6^+_1$     & NA              & 0.4  & 7.2 \\
             & $(10^+_1) \rightarrow (8^+_1)$  &  30(6)          & 29.0 & 218   \\
             & $(12^+_1) \rightarrow (10^+_1)$ &24.4$^{23}_{20}$ & 26.2 & 193 \\
             & $6^+_2 \rightarrow 4^+_1$       &  NA             & 17.6 & -\\
             & $8^+_1 \rightarrow 6^+_2$       & NA              & 18.6 & - \\
            & $2^+_2 \rightarrow 2^+_1$        &  NA             &  18.2  &  -\\        
             & $7^-_1 \rightarrow 5^-_1$       &  34.4(3)        & 9.9 & -\\
             & $9^-_1 \rightarrow 7^-_1$       & 33.2(10)        & 19.9 & -\\ 
             & $(11^-_1) \rightarrow 9^-_1$    &  NA             & 32.2 & -\\
\\
$^{196}$Hg   & $2^+_1 \rightarrow 0^+_1$   &  36(7)      & 26.7     & 33.3\\
             & $4^+_1 \rightarrow 2^+_1$   &  20(15)     & 30.1     & 44\\
             & $(6^+_1) \rightarrow 4^+_1$   &  NA         & 33.3   & 47\\
             & $(8^+_1) \rightarrow (6^+_1)$   &  NA         & 23.2 & 43.3\\
             & $(10^+_1) \rightarrow (8^+_1)$  &  34(10)     & 33.3 & 33     \\
             & $(12^+_1) \rightarrow (10^+_1)$ &  37.8(15)   & 33.2 & 35    \\
             & $6^+_2 \rightarrow 4^+_1$       &  NA         & 0.0 & -\\
             & $8^+_1 \rightarrow 6^+_2$       & NA          & 0.1 & - \\
              & $2^+_2 \rightarrow 2^+_1$   &  NA  &  17.8 &  -\\                                     
             & $7^-_1 \rightarrow 5^-_1$   &  33(1)      & 10.6     & -\\ 
             & $9^-_1 \rightarrow 7^-_1$   &  33.6(18)   & 23.6     & -\\
             & $(11^-_1) \rightarrow 9^-_1$  &  NA       & 34.4     & -\\
\\ 
$^{198}$Hg   & $2^+_1 \rightarrow 0^+_1$   &     28(4)     &  28.6  & 27.9   \\
             & $4^+_1 \rightarrow 2^+_1$   &     $>$16     &  39.9  & 37\\
             & $6^+_1 \rightarrow 4^+_1$   &     9.0(8)    &  42.6  & 37\\
             & $8^+_1 \rightarrow 6^+_1$   &     2.6(15)   &  9.0   & 30\\
             & $10^+_1 \rightarrow 8^+_1$  &     49        &  40.2  & 17\\
             & $12^+_1 \rightarrow 10^+_1$ &     43.0(14)  &  42.0  & 17\\
             &  $2^+_2 \rightarrow 2^+_1$  &    0.63(8)    &  12.6  &  29 \\ 
             & $6^+_2 \rightarrow 4^+_1$       &  NA       & 0.7 & -\\
             & $8^+_1 \rightarrow 6^+_2$       & NA        & 4.0 & - \\
             & $2^+_2 \rightarrow 2^+_1$       &  0.63(8)       &  12.6 & 29\\                                     
             & $7^-_1 \rightarrow 5^-_1$   &     29.5(5)   &  28.4    & -\\ 
             & $9^-_1 \rightarrow 7^-_1$   &     39(7)     &  31.4    & -\\
             & $11^-_1 \rightarrow 9^-_1$  &     NA        &  40.3    & -\\
\\

$^{200}$Hg   & $2^+_1 \rightarrow 0^+_1$    &  26(2)   &  26.9     & 24.5\\
             & $4^+_1 \rightarrow 2^+_1$    &  20(9)   &  37.5    & 34\\
             & $6^+_1 \rightarrow 4^+_1$    &  46(4)   &  34.4    & 31\\
             & $8^+_1 \rightarrow 6^+_1$    &  41(14)  &  38.1    & 19\\
             & $10^+_1 \rightarrow 8^+_1$   &  NA      &  0.6    & -\\
             & $12^+_1 \rightarrow 10^+_1$  & 36(16)   &  33.0   & -\\
  &  $2^+_2 \rightarrow 2^+_1$ &  1.0(8) &  7.3     &  8.4 \\
             & $2^+_2 \rightarrow 0^+_1$ &  0.1(8) &  0.3     & 0.36\\
             
             & $7^-_1 \rightarrow 5^-_1$     &  $>$13     &  13.2   & - \\
             & $9^-_1 \rightarrow 7^-_1$     &  25.1(10)  &  30.6  & -  \\
             & $11^-_1 \rightarrow 9^-_1$    &  NA        &  35.6  & -  \\

\hline
\hline
 \label{t_be2Even}
\end{tabular}
\end{table*}

In $^{198}$Hg, the calculated reduced $E2$ strength between 2$^+_1$ and 0$^+_1$ states is in excellent agreement with the experimental data. Large discrepancies between the calculated and measured values are observed for the $E2$ transitions $6^+_1 \rightarrow 4^+_1$ and $8^+_1 \rightarrow 6^+_1$, both of which are much smaller than the other yrast transitions.
Since such hindered $B(E2)$ values are obtained for the transitions associated with the $6^+_2$ state, the $6^+_1$ and the $6^+_2$ levels may be interchanged. We will revisit the competition of the $6^+$ levels later.

In $^{200}$Hg, the reduced transition probability between 2$^+_1$ and 0$^+_1$ is 26.9 W.u., which is very close to the experimental value, 26(2) W.u. The $E2$ strength observed for the transition $6^+_1 \rightarrow 4^+_1$ is reduced compared to the experimental data. The $E2$ strength between 8$^+_1$ and 6$^+_1$ is measured with large uncertainty, and our calculated value lies within it. 
The calculated transition strength between 8$^+_1$ and 10$^+_1$ states is very small, indicating the sharp change of dominant configurations.  
The 8$^+_1$ state is a member of the ground-state band in $^{200}$Hg where the $\nu 0i_{13/2}$ orbit is almost completely filled, and the 10$^+_1$ and 12$^+_1$ states appear due to two-quasineutron excitation from the $\nu 0i_{13/2}$ orbit. 

\begin{figure*}[htbp] 
    \includegraphics[width=75mm,clip]{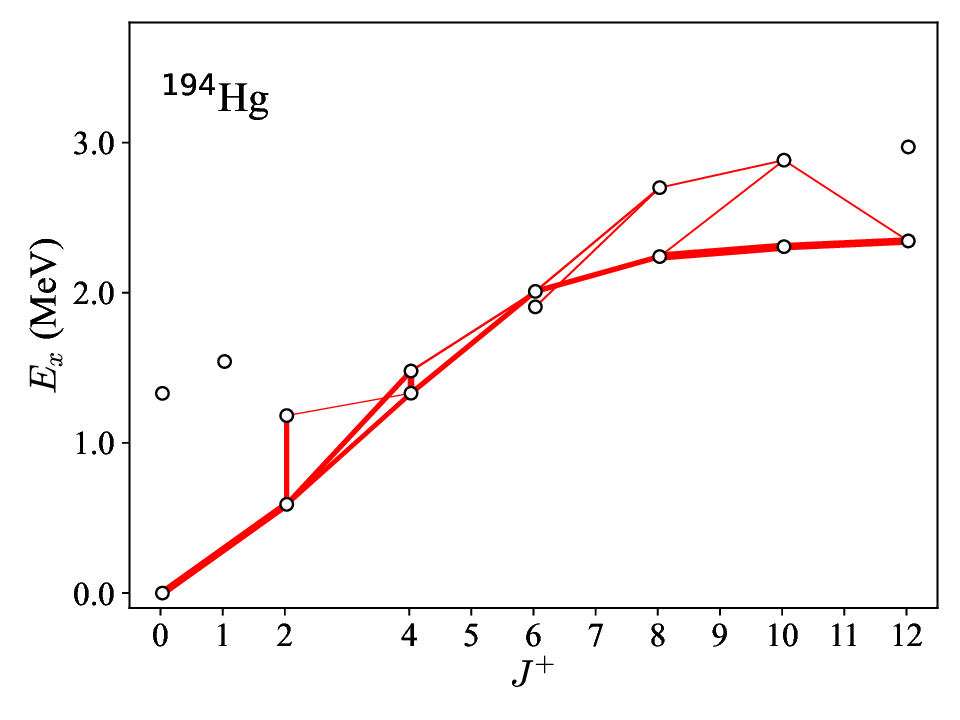}    
    \includegraphics[width=75mm,clip]{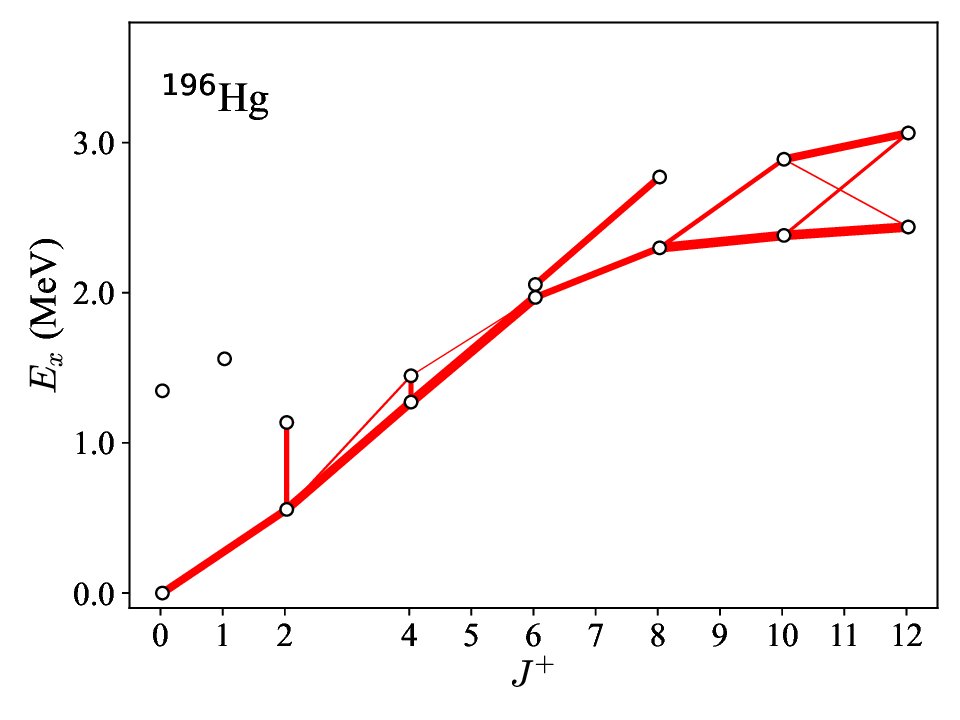}
    \includegraphics[width=75mm,clip]{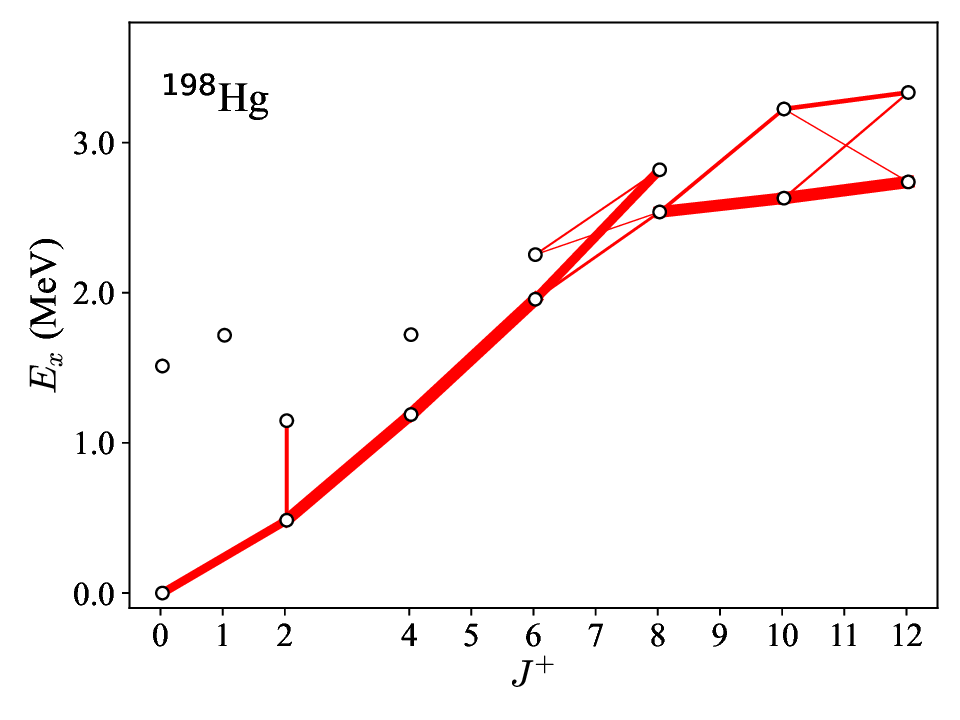}  
     \includegraphics[width=75mm,clip]{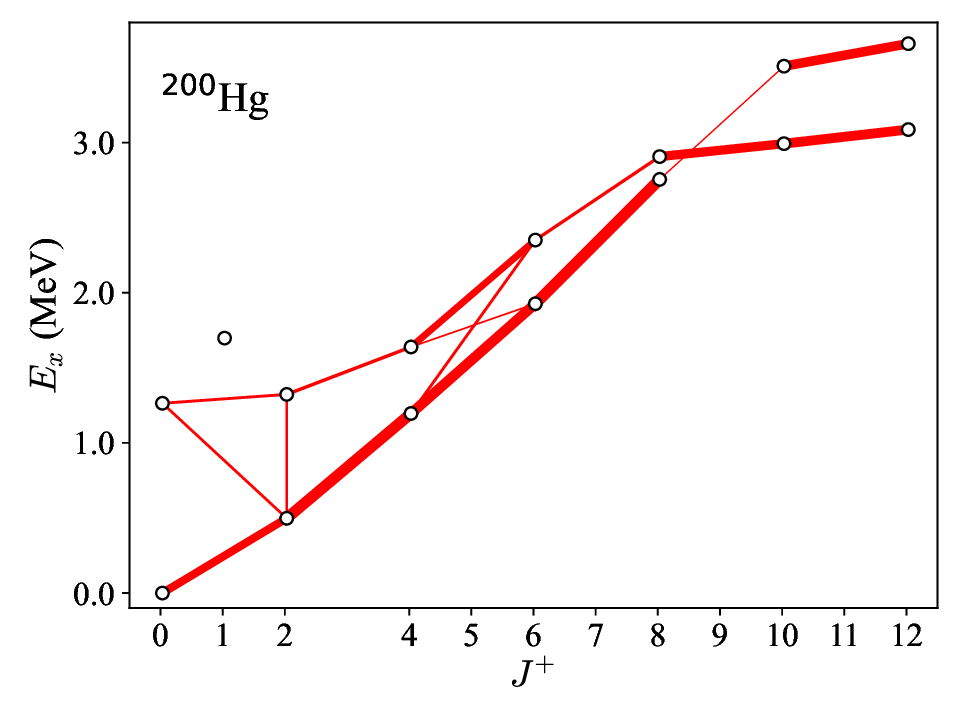}
\caption{\label{fig:E2-evenHg} $E2$ maps for $^{194}$Hg (upper left panel), $^{196}$Hg (upper right panel), $^{198}$Hg (lower left panel) and $^{200}$Hg (lower right panel).
Shell-model excitation energies of even-parity states are plotted as circles against their spins. 
Widths of the red solid lines are proportional to the $B(E2)$ transition strengths. Only the B(E2) values larger than 200 $e^2$fm$^4$ are shown.
}
\end{figure*}

To visualize the E2 probabilities, Figure \ref{fig:E2-evenHg} shows the shell-model $E2$ transition strengths for the positive-parity states of Hg isotopes. 
In $^{194}$Hg, the yrast levels are connected by the strong  $B(E2)$ transitions except that the $6^+_2$ state belongs to the ground-state band. 
In $^{196}$Hg, the yrast $0^+_1$,  $2^+_1$,  $4^+_1$,  $6^+_1$,  $8^+_1$,  $10^+_1$, and  $12^+_1$ are connected without exception. 
In $^{198}$Hg and  $^{200}$Hg, the ground-state bands seem disconnected at the $8^+$ levels due to the sudden structural changes.
These characteristic behaviors will be discussed further in Subsect.~\ref{subsec3_1_3}.

The ratio $B_{4/2}= B(E2; 4^+_1 \rightarrow 2^+_1)/B(E2; 2^+_1 \rightarrow 0^+_1)$ plays an important role in understanding the collective behavior of nuclei across the nuclear chart. The $B_{4/2}$ ratio is exactly 2 in the harmonic vibrator model, while it has a value of $\sim 1.43$ in the case of an ideal rotor. In almost all collective models, the  $B_{4/2}$ ratio is always greater than 1, and can have a value close to one or less than one only in case of nuclei conserving seniority quantum number (observed in semi-magic nuclei) \cite{BE2_ratio1}. The anomalous behavior of observed $B_{4/2}$ ($B_{4/2} < 1$) found in various nuclei, which cannot be explained from any existing collective models, are extensively discussed by Cakirli \textit{et al.} in \cite{BE2_ratio1}. The $B_{4/2}$ ratios are then remeasured for some nuclei in \cite{BE2_ratio2, BE2_ratio3} where it is found that they have a value larger than 1. The interacting boson model in \cite{sh_coIBMCM, Hgdata2019, Hgdata2018} cannot explain the anomalous nature of $B_{4/2}$. The $B_{4/2}$ ratio in the Hg nuclei (Table \ref{t_be2Even}) was measured by Olaizola \textit{et al.} in \cite{Hgdata2019} where it is claimed that the ratios have values above one within experimental uncertainty. In the shell-model result, the calculated $B_{4/2}$ ratios are also larger than 1 and lie within $1.1 \le B_{4/2} \le 1.4$ for $^{196-200}$Hg. On the other hand, this ratio is only 0.63 for $^{194}$Hg, suggesting a configuration change in going from $J=2$ to 4.

Theoretical studies on the energy systematics, transition rates, and moments of even-even Hg isotopes have been carried out using the IBM framework in \cite{sh_coIBMCM, Hgdata2019, Hgdata2018, IBM2}. The two IBM approaches that have been used to study these isotopes are IBM-2 \cite{IBM2} and IBM-CM (Configuration mixing) \cite{sh_coIBMCM}. In the former, the parameters of the Hamiltonian are derived from self-consistent mean field calculation using the microscopic Gogny-D1M energy density functional. In contrast, the latter's parameters are fixed from the experimental energies and $B(E2)$ values. One should refer to those articles (Ref. \cite{sh_coIBMCM} and Ref. \cite{IBM2}) to find a detailed description and the other differences between these two approaches. The observed $E2$ strengths for the transitions $2^+_1 \rightarrow 0^+_1$ and $4^+_1 \rightarrow 2^+_1$, were compared with the IBM results (IBM-2 and IBM-CM) in \cite{Hgdata2019, Hgdata2018}. Olaizola and collaborators found a better agreement of the newly measured $B(E2)$ values with IBM-CM calculated values \cite{Hgdata2019}. Here, we have also compared the shell-model results with those obtained from IBM-CM in Table \ref{t_be2Even}. The IBM-CM results are available for transition between positive parity states and are extracted from tables and graphs of Ref. \cite{sh_coIBMCM}. The $E2$ transition strengths between high spin states (for $J^{\pi} \ge 6$), evaluated in the IBM framework, suffer significant deviation from the experimental value, while the shell-model results are close to it. The collective components contributing to $E2$ strengths are well captured by the shell-model wave functions and on average the shell model provides better results for the $E2$ transition strengths.

\begin{table}
\centering
\caption{Comparison of calculated and experimental \cite{NNDC, mu_data_table, iaea} quadrupole ($Q$) and magnetic moments ($\mu$) of even-Hg isotopes in the units of $e$b and $\mu_N$, respectively.}

\begin{tabular}{ lcccccccccc } 
 \hline
 \hline
& \hspace{2cm}   &  \multicolumn{2}{c}{$Q$ ($e$b)} &  & \multicolumn{2}{c}{$\mu$ ($\mu_{N}$) }   \\
 \cline{3-4}
 \cline{6-7}
Nucleus &$J^{\pi}$&  Expt. & SM & \hspace{1cm} & Expt. & SM \\
 \hline
\\
 $^{194}$Hg   & 2$^+_1$  &  NA       &  0.61    & &  NA        & 0.82    \\
  	          & 4$^+_1$  &  NA       &  -0.52    & &  NA        & 0.30    \\
 	          & 7$^-_1$  &  NA       & 0.98     & & NA         &-0.46         \\
 	          & 9$^-_1$  &  NA       & 0.84     & & NA         & 0.13     \\
 	          & 10$^+_1$ &  NA       & 1.17     & & -2.4(4)    & -1.17       \\
 	          & 12$^+_1$ &  NA       & 1.51     & & -2.9(4)    & -1.21       \\
\\
 $^{196}$Hg   & 2$^+_1$  &  NA       &  0.66    & & NA         & 0.63        \\ 
  	          & 4$^+_1$  &  NA       &  0.26    & & NA         & 0.91         \\
 	          & 7$^-_1$  &  NA       &  1.01    & & -0.21(12)  & -0.36 \\
 	          & 9$^-_1$  &  NA       &  1.07    & & NA         & 0.19     \\
 	          & 10$^+_1$ &  NA       &  1.32    & & -1.9(6)    & -1.15        \\
 	          & 12$^+_1$ &  NA       &  1.65    & & -2.3(7)    & -1.16        \\
 \\
 $^{198}$Hg     & 2$^+_1$  &  0.68(12)   &  0.77    & & 0.76(6)    & 0.71         \\ 
                & 4$^+_1$  &  NA         &  0.87    & & 1.6(2)     & 0.93         \\
 	          & 7$^-_1$  &  NA       &  1.44    & & -0.23(10)  & -0.01         \\
 	          & 9$^-_1$  &  NA       &  1.47    & & NA         & 0.25      \\
 	          & 10$^+_1$ &  NA       &  1.37    & & -1.8(8)    & -1.14        \\
 	          & 12$^+_1$ &  NA       &  1.57    & & -2.2(10)   & -1.14        \\
\\
$^{200}$Hg    & 2$^+_1$  &  0.96(11) &  0.82    & &  0.65(5)   & 0.71       \\
 	          & 4$^+_1$  &  NA       &  1.09    & &  1.02(17)  & 0.95       \\
 	          & 7$^-_1$  &  NA       &  1.26    & &  NA        & 0.36        \\
 	          & 9$^-_1$  &  NA       &  1.32    & &  NA        & 0.32        \\
 	          & 10$^+_1$ &  NA       &  1.26    & &  NA        & -1.09      \\
 	          & 12$^+_1$ &  NA       &  1.49    & &  NA        & -1.05    \\
 \\
 \hline
 \hline
 \label{t_qmEven}
\end{tabular}
\end{table}

\begin{figure}[h]

\centering
\includegraphics[width=80mm]{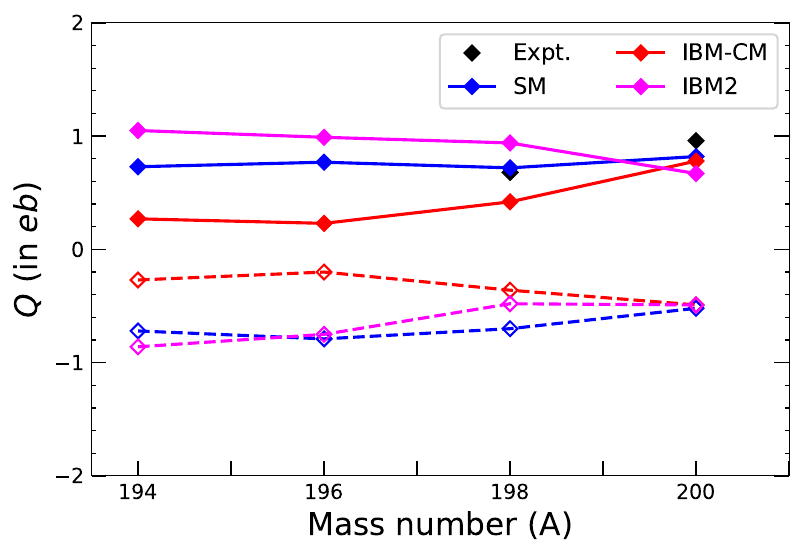}
\caption{Quadrupole moments of $2_1^+$ (filled symbols) and $2_2^+$ (open symbols) states from different models.}\label{Q_2values_EvenHg}
\end{figure}

The shell-model results for spectroscopic quadrupole moments ($Q$) of the lowest $2^+_1$, $4^+_1$ and isomeric states ( $7^-_1$, $9^-_1$, $10^+_1$ and $12^+_1$ )  \cite{Hgdata2021} of the even-$A$ Hg isotopes are listed in Table \ref{t_qmEven}. The observed $Q$ values of $2^+_1$ states in $^{198,200}$Hg are 0.68(12) $e$b and 0.96(11) $e$b, respectively, while the corresponding calculated values are 0.77 $e$b and 0.82 $e$b, respectively. The observed $Q$ moments are well reproduced by the shell model. No other experimental data are available for quadrupole moments in this region, while we have provided the shell-model predictions. The quadrupole moments of the lowest two $2^+$ states have been calculated in the IBM framework in Refs. \cite{sh_coIBMCM} and \cite{IBM2}. Those calculated values (extracted from graphs) are compared with shell-model results and available experimental data in Fig. \ref{Q_2values_EvenHg}. Both the IBM and the shell-model studies well reproduce the observed $Q(2_1^+)$ value in $^{200}$Hg. In the case of $^{198}$Hg, the shell-model calculated value is closer to the experimental data than IBM results. The calculated $Q$ values ($Q(2_1^+)>0$ and $Q(2_2^+)<0$) by these models indicate the existence of an oblate ground state band in the concerned even Hg isotopes. The shell-model results show that the isomeric states have larger $Q$ values than those of $2^+_1$ and $4^+_1$ states in the corresponding Hg isotopes. 

\begin{figure} 
    \includegraphics[width=80mm,height=55mm]{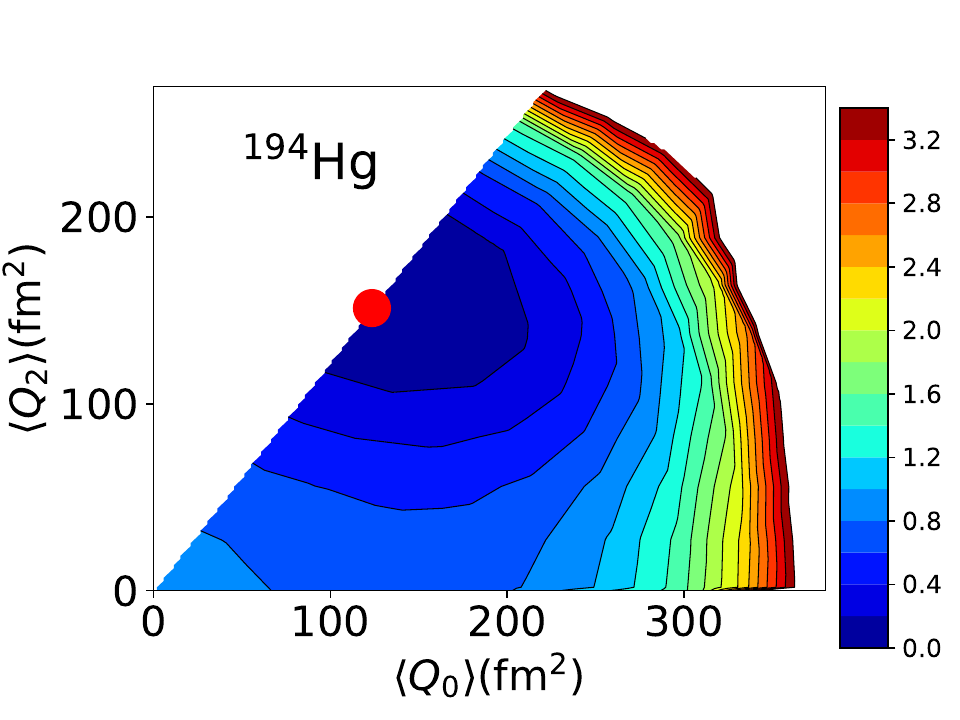}
    \includegraphics[width=80mm,height=55mm]{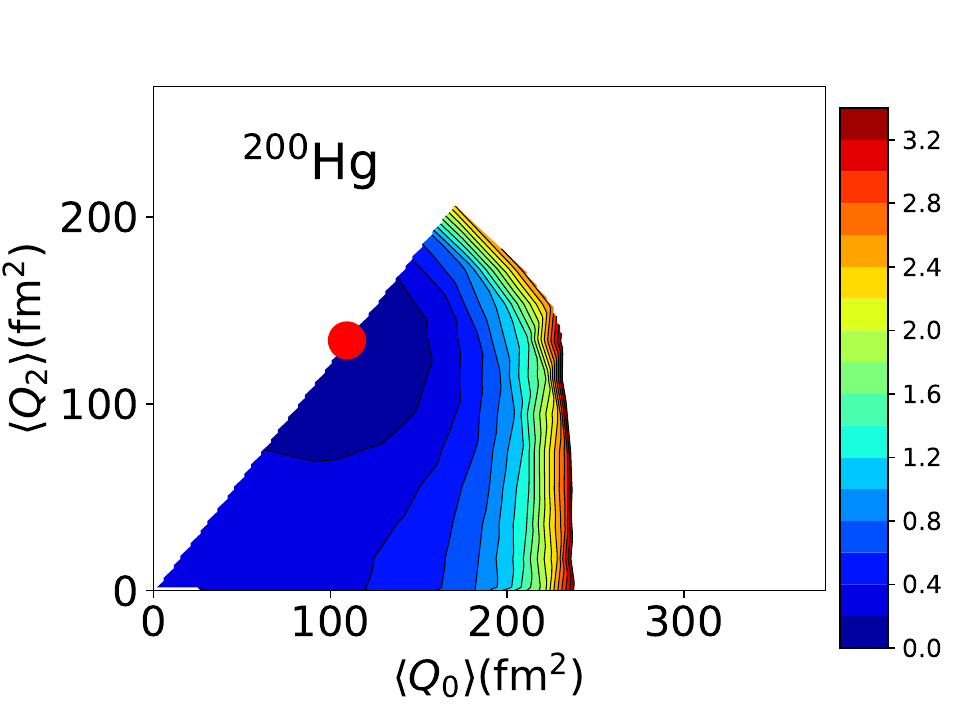}
    \caption{
    \label{fig:pes}
    Energy surfaces of $^{194, 200}$Hg drawn by the Q-constrained HFB calculations. The red symbols denote the minimum points.}
\end{figure}

To discuss the deformation further, we present an energy surface of $^{194}$Hg and $^{200}$Hg drawn by the Q-constrained Hartree-Fock-Bogoliubov (HFB) calculations employing the shell-model interaction in Fig.~\ref{fig:pes}. The energy is evaluated as a function of the quadrupole deformation $\langle Q_0 \rangle $ and $\langle Q_2 \rangle$ with  $\langle Q_m \rangle = \langle r^2Y^{(2)}_{m} \rangle$.
The Q-constrained HFB wave functions are obtained with the variation after the particle-number projection using the code \cite{ShimizuCode,NShimizu2021}. 
The energy surface of $^{194}$Hg presents the oblate deformation, which is consistent with the positive quadrupole moment of the $2^+_1$ state. Similar shapes are seen also in $^{196,198}$Hg though these surfaces are not shown. 
In the $^{200}$Hg case, the surface is rather flat in the direction of $\beta$ near the spherical shape, which may indicate the surface vibrational mode, one of whose feature is the approximate degeneracy of the $0^+_2, 2^+_2, 4^+_1$ states as shown in Fig.~\ref{fig:200Hg}.
The Gogny density functional also gives modest oblate deformation \cite{IBM2}.

The magnetic moments ($\mu$) of the even-even Hg isotopes are also compared with their corresponding experimental data in Table \ref{t_qmEven}. 
The $\mu$ values of $2^+_1$ state for $^{198,200}$Hg obtained by the shell model are in fair agreement with the experimental data. The $\mu$ values for these states are predicted for $^{194,196}$Hg where experimental data are unavailable. The $g$-factor of a state with spin $I$ is connected to the magnetic moment as $\mu = gI \mu_N$. If the magnetic moments of a rotational band are created by the orbital angular momentum alone, their $g$ factors have a state-independent value of $Z/A$ by assuming that the proton and neutron $g_l$ values are 1 and 0, respectively. The shell-model calculations show that the $g$ factors for the $4^+_1$ states are smaller than those of the $2^+_1$ states, thus indicating gradual structure changes with increasing spin. For the $10_1^+$ and $12_1^+$ states, the calculated magnetic moments are all negative and remain almost constant along the chain. These negative magnetic moments, which are obtained by experiment \cite{gfactor_196_198Hg, gfactor_194_196Hg}, are a clear signature of the dominance of neutron configurations of the $j_>$ orbital because the spin $g$ factor of a neutron is negative. The calculated magnetic moments have smaller magnitudes than the measured values, although there are large experimental uncertainties. Note that these magnetic moments are sensitive to the choice of $g_l$ of a neutron. In the shell-model results, the $7_1^-$ state is associated with a very small $\mu$ value in the case of $^{198}$Hg. 
 
\subsubsection{\textbf{Evolution of configuration}\\}
\label{subsec3_1_3}
As discussed in Sect.~\ref{subsec3_1_2}, the calculated and the experimental $E2$ matrix elements and magnetic moments indicate that the dominant configurations change along the yrast $0^+$ to $12^+$ levels. In this section, we survey in more detail how these configuration changes occur on the basis of the analysis of shell-model wave functions.

For this purpose, it is helpful to examine how the total angular momentum $J_z$ is carried by each orbital $\alpha$. More specifically, we decompose the operator $\hat{J_z}$ into each orbital's contribution: 
\begin{equation}
    \hat{J_z} = \sum_{\alpha j} \hat{J_z}(\alpha j), 
\end{equation}
where $\hat{J_z}(\alpha j)$ is defined as 
\begin{equation}
    \hat{J_z}(\alpha j) = \sum_m m a^{\dag}_{\alpha j m}a_{\alpha j m}, 
\end{equation}

with $j$ and $m$ being the total and the $z$-component of the angular momentum of the single-particle state, respectively. 
By using this operator, the fraction of $\langle \hat{J_z} (\alpha j) \rangle$ in the total $\langle \hat{J_z} \rangle$ is expressed as 
\begin{equation}
    R_{\alpha j} \equiv \frac{\langle \hat{J_z} (\alpha j) \rangle}{\langle \hat{J_z} \rangle} ,
\end{equation}
where $\langle \rangle$ is the expectation value of a state of interest. The $R_{\alpha j}$ values spread over various orbitals for a rotational state, whereas they are concentrated in one or a few orbitals if particular configurations dominate a state.

\begin{figure}[t] 
\centering
    \includegraphics[width=90mm,clip]{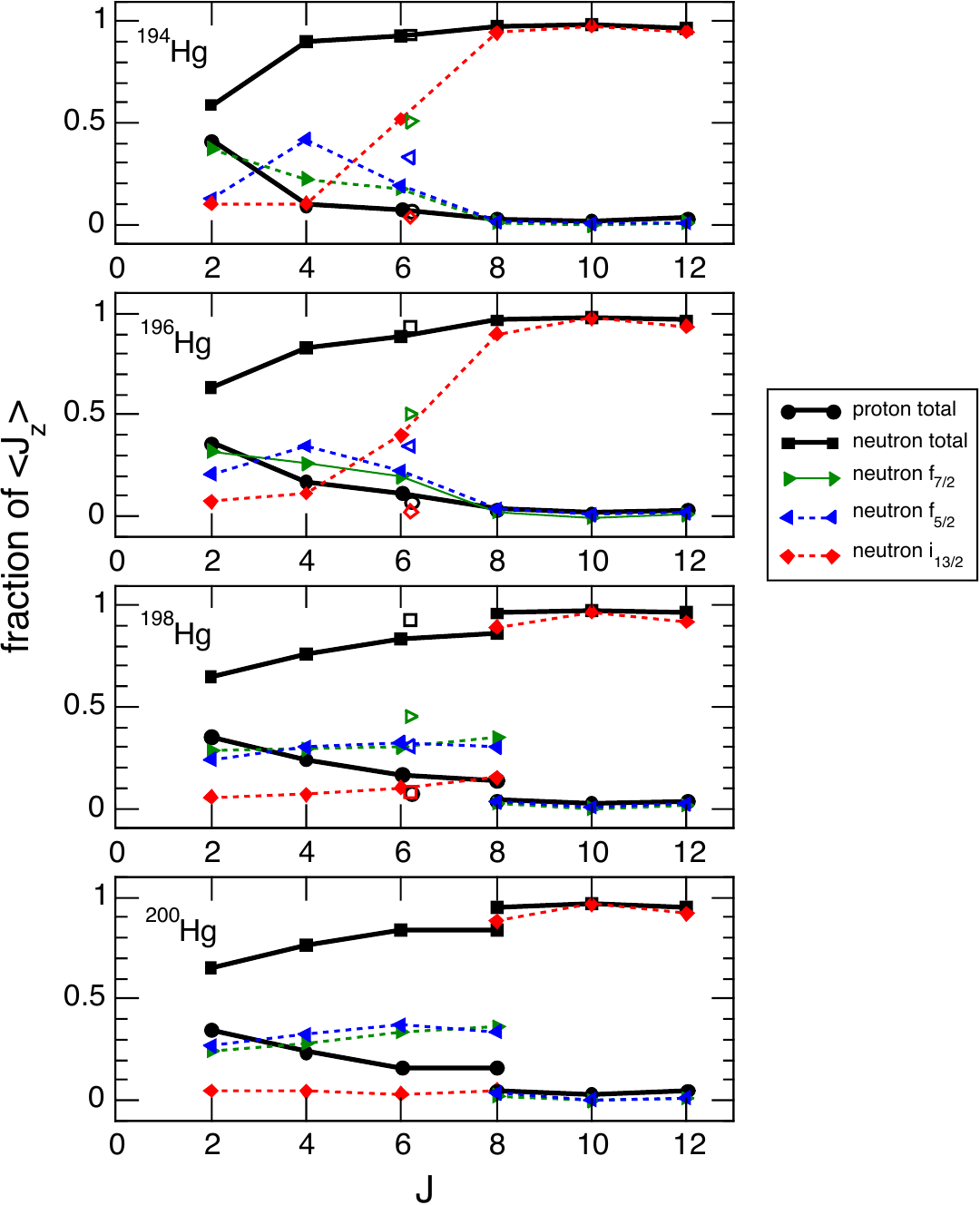}
	\caption{ Fraction of $\langle \hat{J_z} (\alpha j) \rangle$ for $2^+$, $4^+$, $6^+$, $8^+$, $10^+$ and  $12^+$ states in $^{194,196,198,200}$Hg. The lines connect the states having large $E2$ matrix elements: the $2^+_1$, $4^+_1$, $6^+_2$, $8^+_1$, $10^+_1$ and $12^+_1$ states in $^{194}$Hg, the $2^+_1$, $4^+_1$, $6^+_1$, $8^+_1$, $10^+_1$ and $12^+_1$ states in $^{196}$Hg, the $2^+_1$, $4^+_1$, $6^+_1$, and $8^+_2$ states and the $8^+_1$, $10^+_1$ and $12^+_1$ states in $^{198}$Hg, and the $2^+_1$, $4^+_1$, $6^+_1$, and $8^+_1$ states and the $8^+_2$, $10^+_1$ and $12^+_1$ states in $^{200}$Hg. The unfilled symbols at $J=6$ in $^{194,196,198}$Hg stand for the ``intruder" $6^+$ states ($6^+_1$ in $^{194}$Hg and $6^+_2$ in $^{196,198}$Hg).
 }
 \label{fig:j_fraction}
\end{figure}

The calculated $R_{\alpha j}$ values are summarized in Fig.~\ref{fig:j_fraction}. We start discussions with $^{200}$Hg, whose situation is simpler than in the case of the other nuclei. For the $2^+_1$ state (at $J=2$), both the protons and the neutrons almost equally contribute to $\langle \hat{J_z} \rangle$, suggesting a collective behavior. The $4^+_1$, $6^+_1$, and $8^+_1$ states have similar distributions of $R_{\alpha j}$, but with a gradual change the neutron contributions become more dominant. 
The measured $g$ factors of the $2^+_1$ and the $4^+_1$ states, $0.32(3)$ and $0.26(4)$, respectively, show a slight decrease with spin and may point to such a gradual change. These configuration changes, however, occur rather smoothly, and hence the calculated $E2$ matrix elements between those members are kept large. When $J$ is moved to $10$, the $R_{\alpha j}$ values are concentrated in the neutron $i_{13/2}$ orbital and this tendency persists in the $12^+_1$ state. The $8^+$ member of such a character appears as the $8^+_2$ state. As a result, the calculated $B(E2; 10^+_1 \to 8^+_1)$ value becomes small as shown in Table~\ref{t_be2Even} and Fig.~\ref{fig:E2-evenHg} due to the sharp change of configuration. 

In $^{198}$Hg, the distributions of $R_{\alpha j}$ are similar to those of $^{200}$Hg but with the $8^+_1$ and the $8^+_2$ states interchanged. Namely, the $8^+_1$ state ($8^+_2$) has a $R_{\alpha j}$ distribution close to those of $10^+_1$ and $12^+_1$ states ($2^+_1$, $4^+_1$ and $6^+_1$ states). This interchange occurs because the $i_{13/2}$ band is lowered (see Fig.~\ref{fig:EvenA_posSym}) in this nucleus because the neutron Fermi surface goes down and becomes closer to the $i_{13/2}$ orbital, which is almost completely filled in the ground state. Thus, along the yrast line, a sharp configuration change occurs in going from $6^+_1$ to $8^+_1$. However, if the $6^+_1$ state is a member of the ground band, the $B(E2;6^+_1 \to 4^+_1)$ should be rather large as obtained in the calculation. This is in contrast to the experimental data as shown in Table~\ref{t_be2Even}. A possible scenario to account for the observed small $B(E2)$ values for the $8^+_1 \to 6^+_1$ and the $6^+_1 \to 4^+_1$ transitions (see Table~\ref{t_be2Even}) would be the appearance of the $6^+_1$ state that is dominated by different configurations from the $4^+_1$ state as well as from the $8^+_1$ state. 
A candidate for such an ``intruder" state is the calculated $6^+_2$ state, whose $R_{\alpha j}$ distribution is plotted by unfilled symbols in Fig.~\ref{fig:j_fraction}. Its $\langle \hat{J_z} \rangle$ value is contributed almost solely by the neutron $f_{7/2}$ and the $f_{5/2}$ orbitals. Since the calculated $R_{f_{7/2}}$ value to the $R_{f_{5/2}}$ value, 1.47, is close to that of the $(f_{7/2})^{-1}(f_{5/2})^{+1}$ configuration, 1.4($=\frac{7/2}{5/2}$), this state is considered to be dominated by the $(f_{7/2})^{-1}(f_{5/2})^{+1}$ $1p$-$1h$ state.

In $^{196}$Hg, all the yrast states of $0^+_1$, $2^+_1$, $4^+_1$, $6^+_1$, $8^+_1$, $10^+_1$ and $12^+_1$ are calculated to be connected with large $E2$ matrix elements as presented in Table~\ref{t_be2Even} and Fig.~\ref{fig:E2-evenHg}. This looks puzzling when one accepts the fact that the dominant configurations of the high-spin members are completely different from those of the low-spin members. The key to understanding this apparently contradicting result is a gradual configuration change with changing spin. As presented in Fig.~\ref{fig:j_fraction}, similar to $^{198}$Hg, the $8^+_1$, $10^+_1$ and $12^+_1$ members in $^{196}$Hg are dominated by the almost pure $i_{13/2}$ band. In contrast, the $R_{\alpha j}$ distributions in the $6^+_1$ state indicate that the $i_{13/2}$ configuration strongly admixes with the ground-band configuration there. This results in a smooth transition of these two bands. Similar to the case of $^{198}$Hg, the $(f_{7/2})^{-1}(f_{5/2})^{+1}$ $1p$-$1h$ state appears as the $6^+_2$ state in $^{196}$Hg. This state is located only 87~keV above the $6^+_1$ state, thus pointing to a severe competition of these two states. 

The situation of $^{194}$Hg is more or less similar to $^{196}$Hg, but there are two noticeable differences. First, the $(f_{7/2})^{-1}(f_{5/2})^{+1}$ $1p$-$1h$ state appears as the yrast state with 102~keV lower than the $6^+$ member of the ground band. Second, the difference between the $2^+_1$ and the $4^+_1$ states in the $R_{\alpha j}$ distributions is enlarged. The calculated relatively small $B(E2; 4^+_1 \to 2^+_1)$ value is thus caused.   

To briefly summarize this section, the shell-model calculations predict subtle configuration changes and/or mixing with changing spin, which have different characteristics from nucleus to nucleus. Part of the calculated results are supported by experiments, while others are yet to be confirmed, thus motivating future experiments.

\subsection{\textbf{Odd Hg isotopes}}
This section contains the shell-model results for odd-$A$ Hg isotopes. The excitation spectra are presented in Sec. \ref{subsec3_2_1} while the $E2$ transition probabilities, quadrupole and magnetic moments of the odd-odd Hg are discussed in Sec. \ref{subsec3_2_2}.


\subsubsection{\textbf{Energy spectra}\\}
\label{subsec3_2_1}

In $^{193}$Hg, the parities of the ground state and the first excited state have not yet been confirmed. The shell model predicts the same parity for these states, as suggested by experiments. The states with spins ${3/2}$, ${5/2}$, and ${1/2}$ lie close to each other in the experimental data. In the shell-model results, the ${3/2}^-$ state lies adjacent to the ${5/2}^-$ state, but the ${1/2}^-$ appeared at higher energy than the measured value. 
We obtained a perfect agreement of the order of the positive and negative parity high spin states from the experiment and shell-model theory except ${23/2}^-$. In the shell-model result, the ${23/2}^-$ state is found at lower energy and below the ${21/2}^-$ state. Similar to $^{193}$Hg, the ${3/2}^-$ and ${5/2}^-$ states are closely spaced, and the ${1/2}^-$ state is obtained at higher energy in the calculated spectra of $^{195}$Hg. Multiple spins (${3/2}^-$, ${5/2}^-$, ${7/2}^-$) have been proposed for the state at an energy of 0.410 MeV in the experiment. 
In the shell-model result, the ${7/2}^-$ and ${5/2}^-$ states are closely spaced and observed at $\sim$ 0.445 MeV. 
In $^{195}$Hg, we noticed the same behavior of the ${23/2}^-$ state as observed in $^{193}$Hg. 

\begin{figure}[h] 
\centering
    \includegraphics[width=90mm,height=100mm]{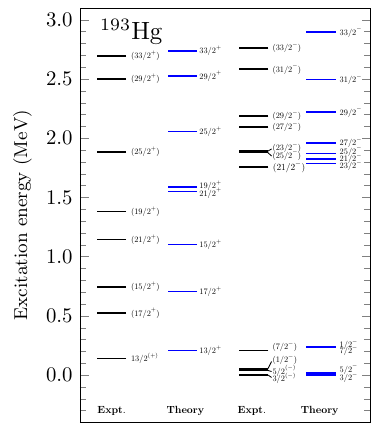}
	\caption{ Comparison between calculated and experimental \cite{NNDC} energy levels for $^{193}$Hg.}
 \label{fig:193Hg}
\end{figure}

\begin{figure}[h] 
\centering
    \includegraphics[width=90mm,height=100mm]{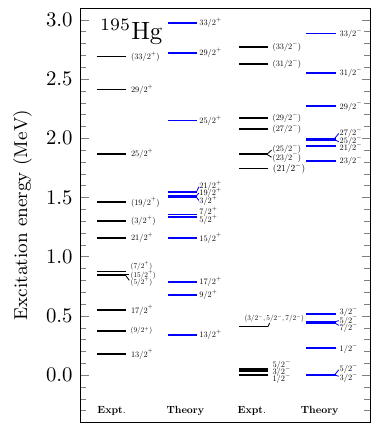}
	\caption{ Comparison between calculated and experimental \cite{NNDC} energy levels for $^{195}$Hg.}
 \label{fig:195Hg}
\end{figure}

\begin{figure}[h] 
\centering
    \includegraphics[width=90mm,height=100mm]{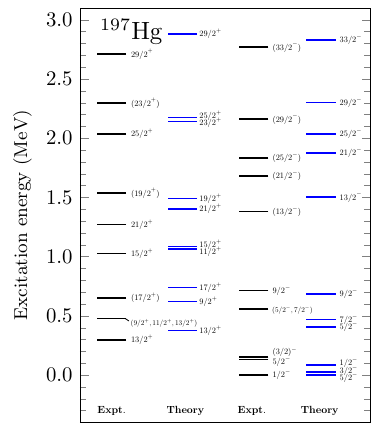}
	\caption{ Comparison between calculated and experimental \cite{NNDC} energy levels for $^{197}$Hg.}
 \label{fig:197Hg}
\end{figure}

\begin{figure}[h] 
\centering
    \includegraphics[width=90mm,height=100mm]{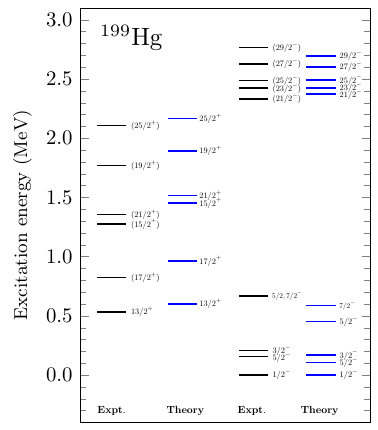}
	\caption{ Comparison between calculated and experimental \cite{NNDC} energy levels for $^{199}$Hg.}
 \label{fig:199Hg}
\end{figure}

The energy spectra of both positive and negative parity states of odd mass Hg isotopes from $A=193$ to $A=199$ are shown in Figs. \ref{fig:193Hg}, \ref{fig:195Hg}, \ref{fig:197Hg} and \ref{fig:199Hg} respectively.

The ${1/2}^-$ ground state appears at an excitation energy of 84 keV in the calculated spectra in $^{197}$Hg. Like experimental data, the ${5/2}^-$ and ${3/2}^-$ states are almost degenerate. The experiment assigns spins ${5/2}^-$ and ${7/2}^-$ for the state situated at an energy of 0.557 MeV. It can be associated with the ${7/2}_1^-$ state in the shell-model result. The order of the high spin negative parity states agrees with experimental data.
Contrast to other odd-$A$ Hg isotopes, order and energies of the lowest ${1/2}^-$, ${5/2}^-$, and ${3/2}^-$ states are well reproduced in the shell-model results in $^{199}$Hg. The experiment assumes the spin ${5/2}^-$ or ${7/2}^-$ for the state observed at energy 0.667 MeV which may correspond to the ${7/2_1}^-$ state in the calculated spectra. The order of high spin positive and negative parity states are in perfect agreement with the experimental data. 

In the case of odd-$A$ Hg isotopes, the shell-model calculations often overestimate the energies of the high spin positive parity states. Although the shell-model calculations well reproduce the order of the high spin negative parity states, the correct order of the ${23/2}^-$ state is not observed in $^{193,195}$Hg. However, the ${23/2}^-$ state is found in the same order as per the experimental data in $^{199}$Hg. We found that though the level scheme of $^{199}$Hg is computed with $E_\textrm{cutoff}$ = 12 MeV, the correct order of the ${23/2}^-$ state is maintained even with lower energy cutoff, $E_\textrm{cutoff}$ = 10 MeV.

The present result was unable to predict the ground-state spin of $^{195}$Hg and its $1/2^-_1$ excitation energy is roughly 200 keV higher than the experimental one.  
In the case of $^{193}$Hg, $^{195}$Hg and $^{197}$Hg, in which we adopted $E_\textrm{cutoff}=10$ MeV, the present study provides us with their $1/2^-_1$ states roughly 200 keV higher than the experimental ones. One of the reasons for this disagreement is the truncation we adopted. 
We found the convergence of the $1/2^-_1$ energy is relatively slower than that of other states, which could be related to the collectivity. 
For example, the ${1/2}^-_1$ excitation energy of $^{197}$Hg decreases by around 60 keV with an increase in $E_\textrm{cutoff}$ from 10 MeV to 11 MeV. 
The $1/2^-_1$ excitation energy of $^{195}$Hg is 0.226 MeV  for $E_\textrm{cutoff}=10$ MeV and 0.146 MeV for $E_\textrm{cutoff}=12$ MeV, closer to the experimental one. 

\begin{figure}[h] 
\centering
    \includegraphics[width=90mm]{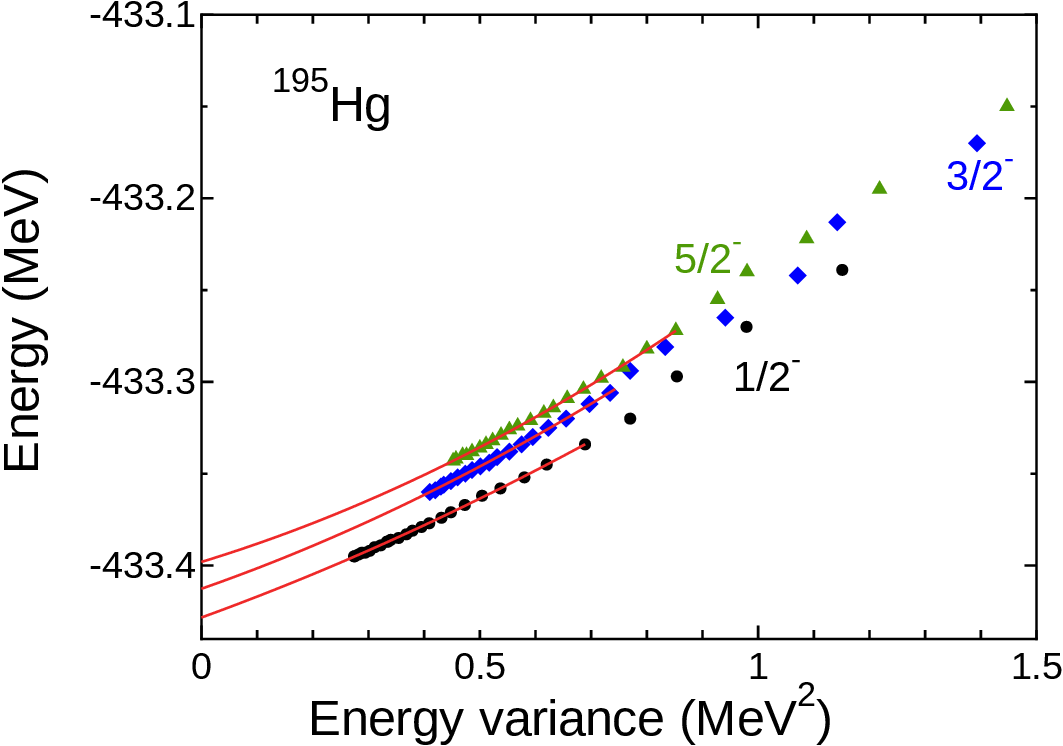}
	\caption{ Energy-variance plot of the QVSM calculation of $^{195}$Hg. The energy expectation values of the QVSM wave functions of the $1/2^-_1$, $3/2^-_1$, and $5/2^-_1$ states are plotted against the corresponding energy variances for each number of the basis states as black circles, blue squares, and green triangles, respectively. These points are $\chi^2$-fitted by the red curves using the second-order polynomials. }
 \label{fig:Hg195_h2}
\end{figure}

To discuss the ground-state spin of $^{195}$Hg employing the KHHE interaction without truncation, we performed the quasiparticle-vacua shell model (QVSM) calculation \cite{qvsm2021}. In the QVSM calculation, the shell-model wave function is approximated as a linear combination of the 24 basis states which are angular-momentum-, parity-, and number-projected quasiparticle vacua provided by the variation after projection. Using this QVSM wave function, we applied the energy-variance extrapolation to estimate the exact shell-model energies precisely.
Figure \ref{fig:Hg195_h2} shows the energy-variance plot for the extrapolation with increasing the number of the basis states from 1 to 24. As the number of the basis states increases the approximation is improved, and the energy and the corresponding energy variance approach the exact shell-model energy and zero, respectively. Therefore, the $y$-intercepts of the fitted curves are the estimation of the exact one by the extrapolation.
The extrapolated results show the $1/2^-_1$ ground state, $3/2^-_1$ ($E_x=$ 16 keV) and $5/2^-_1$ ($E_x=$ 30 keV) excited states.
The ordering of these three states is consistent with the experimental result, unlike the monopole-based truncation. Thus, the KHHE interaction without truncation would predict the correct ground-state spin.

\subsubsection{\textbf{Electromagnetic properties}\\}
\label{subsec3_2_2}

This section follows a similar discussion on the electromagnetic properties for the odd mass Hg isotopes as done in the case of even mass Hg isotopes. Here, the values of the effective charges and $g$- factors for protons and neutrons remain the same as those used in the even-even Hg isotopes.

In odd-mass Hg isotopes, the $E2$ transition strengths have been measured only for transitions between low-lying states. Hence, we have presented the shell-model results for the $E2$ transition probabilities between $1/2_1^-$, $3/2_1^-$, and $5/2_1^-$ states in Table \ref{t_be2odd}. 
The calculated $E2$ strengths for the $5/2^-_1 \rightarrow 1/2^-_1$ and $3/2^-_1 \rightarrow 1/2^-_1$ transitions are in reasonable agreement with experimental data while the $3/2^-_1 \rightarrow 5/2^-_1$ transition of $^{199}$Hg is underestimated. In the shell-model results, the $B(E2;5/2^-_1 \rightarrow 1/2^-_1 )$ is also reduced compared to the experimental value in $^{197}$Hg. The shell model produces small $E2$ strengths for the transitions between low-lying states in the odd-mass Hg isotopes.

\begin{figure*}[htbp] 
    \includegraphics[width=75mm,clip]{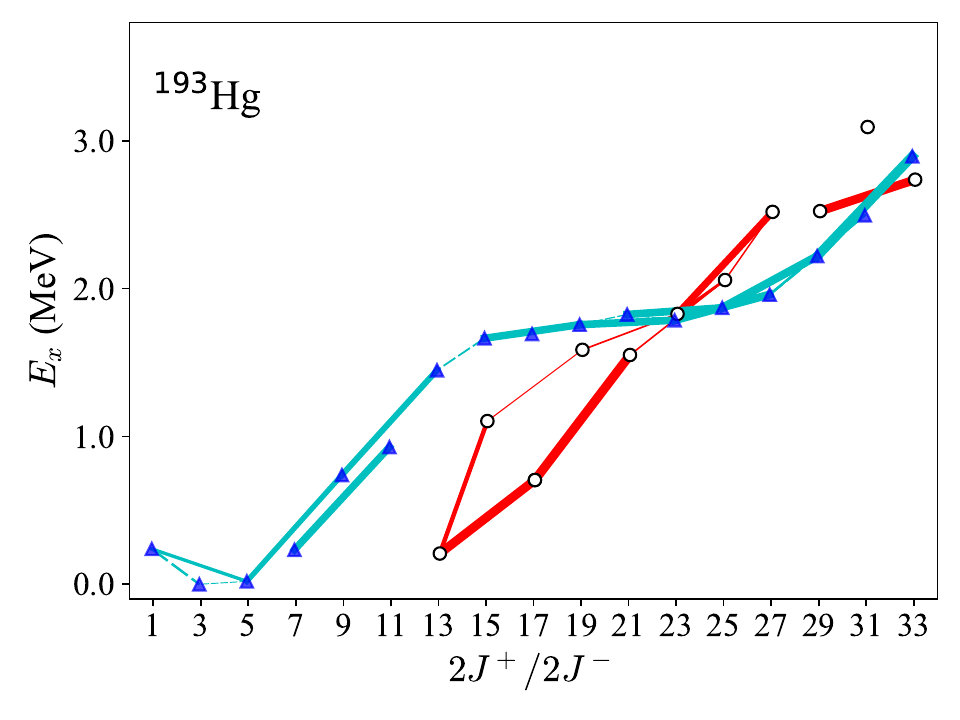}    
    \includegraphics[width=75mm,clip]{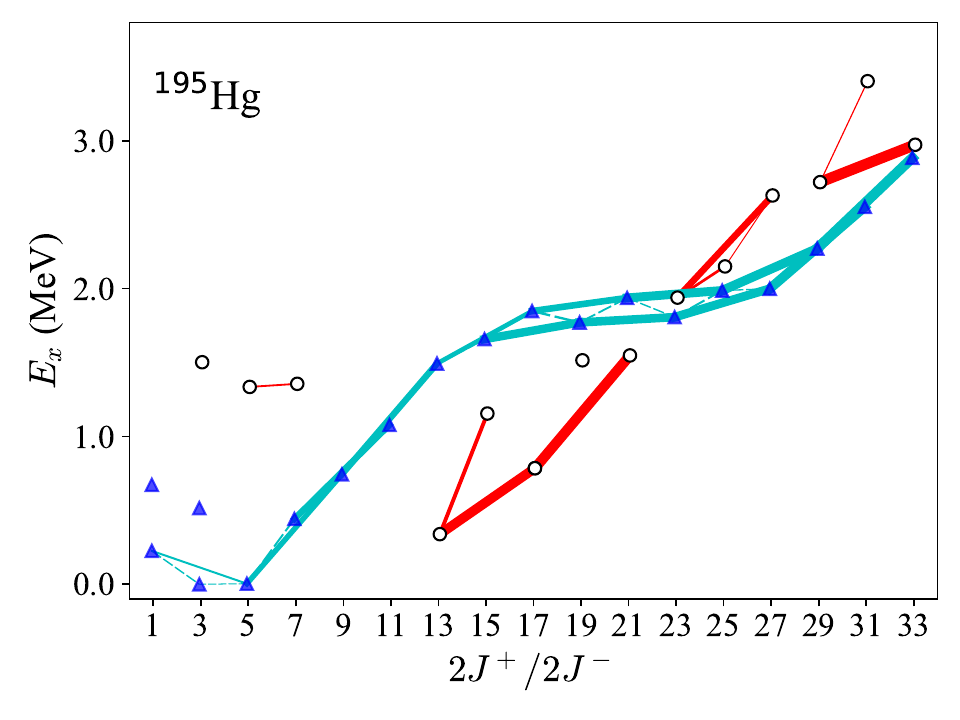}
    \includegraphics[width=75mm,clip]{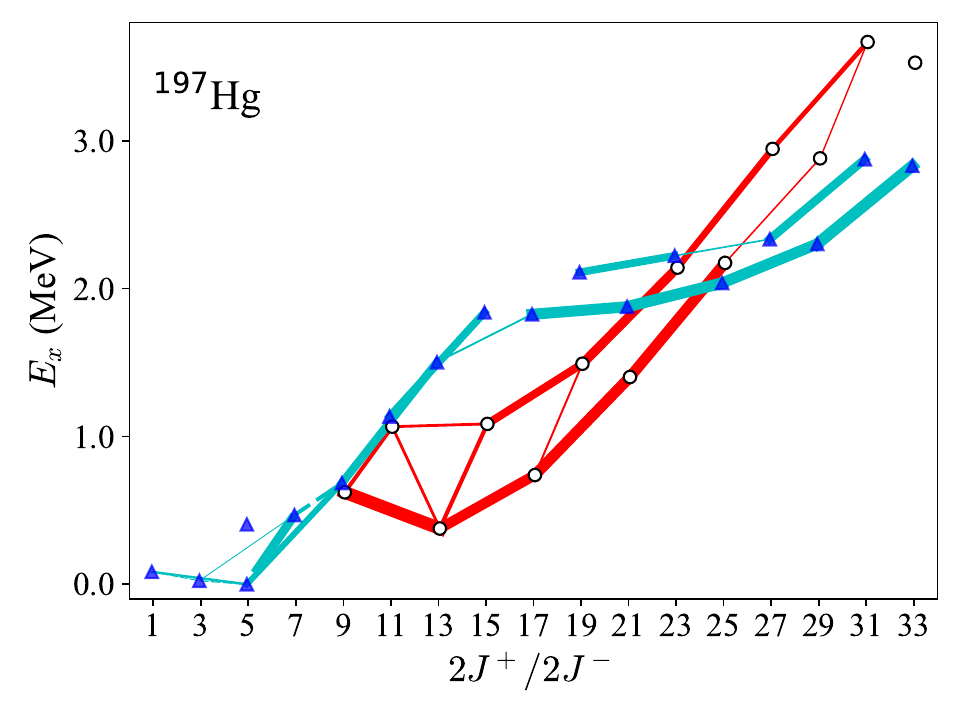}   
    \includegraphics[width=75mm,clip]{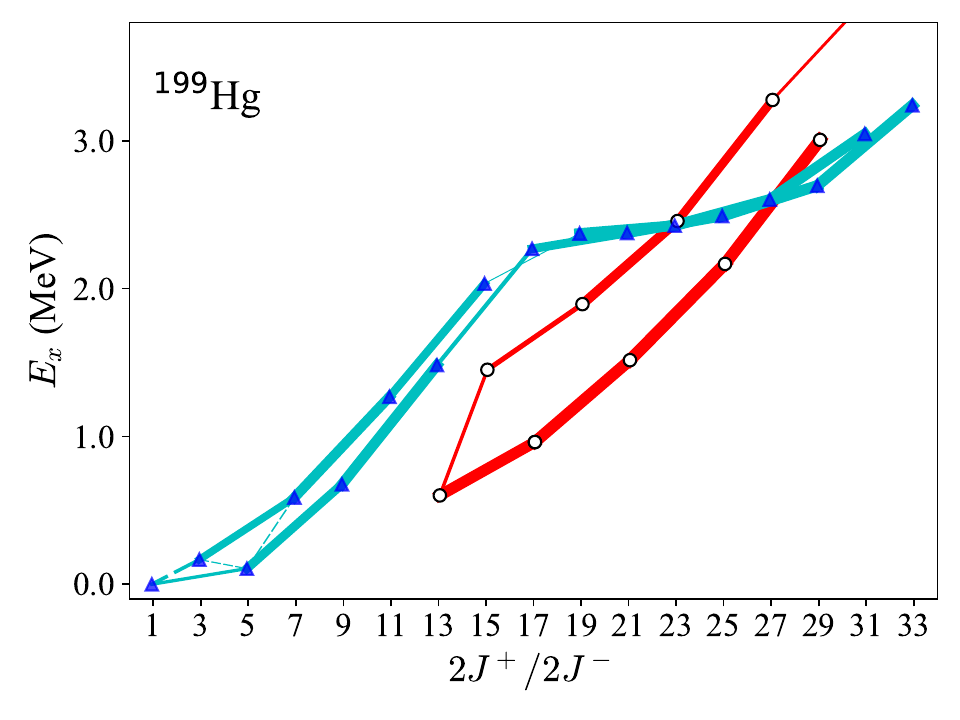}
\caption{\label{fig:E2-oddHg} $E2$ maps for $^{193}$Hg (upper left panel), $^{195}$Hg (upper right panel), $^{197}$Hg (lower left panel) and $^{199}$Hg (lower right panel).
Shell-model excitation energies of positive (negative) parity states are plotted as open black circles (filled blue triangles) against their spins. 
Widths of the red (blue) lines are proportional to the $B(E2)$ transition strengths between positive (negative) parity states. The solid (dashed) lines indicate the $\Delta J=2$ ($\Delta J=1$) transitions. Only the $B(E2)$ values larger than 200 $e^2$fm$^4$ are shown.
}
\end{figure*}

 To visualize the $E2$ transitions and the band structures of the odd-mass Hg isotopes their ``$E2$ maps'' are presented in Fig.~\ref{fig:E2-oddHg}.
The experimental spectra of these odd-mass Hg isotopes show similar common characteristics: a few low-spin negative parity states are observed below $E_x\sim 0.5$ MeV and high-spin negative parity states are observed while only the positive parity states starting from $13/2^+_1$ are seen in the $E_x\sim 1$ MeV region. These features are consistent with the shell-model results and its $E2$ maps indicating that the positive-parity states form a $13/2^+$ band located roughly 1 MeV lower than the corresponding spin negative parity states. Since in most $\gamma$-ray spectroscopy experiments, only the yrast states or its neighboring states are measured, the negative-parity states around $E_x\sim 1$ MeV are screened out by the yrast positive-parity states.
In $^{197}$Hg and $^{199}$Hg, the positive-parity $13/2^+$, $17/2^+$, $21/2^+$ states form a favored band of the $\nu0i_{13/2}$ orbit in the decoupled band and the band starting from $15/2^+$ forms the unfavored band as suggested in Refs. \cite{Proetel-odd-Hg-1974,Hg197_199}.
In contrast, such band structure is contaminated by the configuration mixing in $^{193}$Hg and $^{195}$Hg.

The $E2$ maps indicate that the negative-parity states can be divided into two groups: higher spin states formed by two quasiparticles in the $(\nu0i_{13/2})$ orbit and one quasiparticle in a low-$j$ orbit, and the other groups consisting of only low-$j$ states.
In $^{199}$Hg and $^{197}$Hg, the former group starts from $17/2^-$, and these $E2$ strengths to the other group are relatively small. These states could be interpreted as the coupling of the $(\nu0i_{13/2})$ single-particle state and the ($5^-$, $7^-$) states of the neighboring nuclei suggested in \cite{Proetel-odd-Hg-1974}.
On the other hand, in $^{195}$Hg and $^{193}$Hg, the former groups start from the $13/2^-$ state and they transit rather gradually because the Fermi surface approaches the $(\nu0i_{13/2})$ orbit, which is similar to the case of even-mass isotopes discussed in Subsect.~\ref{subsec3_1_3}.

\begin{table*}
\centering
\caption{Calculated $E2$ strengths (in W.u.) of odd-$A$ Hg isotopes in comparison with experimental values \cite{NNDC}.}
\begin{tabular}{ lccccccc     } 
 \hline
 \hline
 &  \hspace{8cm} & \multicolumn{2}{c}{$B(E2)$}       \\
 \cline{3-4}
Nucleus &  \hspace{0.5cm} $J^{\pi}_i \rightarrow J^{\pi}_f $&  \hspace{0.5cm} Expt. & \hspace{0.5cm} SM \\
 \hline 
 \hline
 
\\
$^{193}$Hg   & $5/2^{(-)}_1 \rightarrow 3/2^{(-)}_1$  &  NA    &  3.4   \\
             & $(1/2^-_1) \rightarrow 5/2^{(-)}_1$    &  NA    &  12.2    \\
             & $(1/2^-_1) \rightarrow 3/2^{(-)}_1$    &  NA    &  9.0    \\

\\
$^{195}$Hg   & $3/2^-_1 \rightarrow 1/2^-_1$    &  $>$60     &  2.5    \\
             & $5/2^-_1 \rightarrow 3/2^-_1$    &  18(10)    &  3.6   \\
             & $5/2^-_1 \rightarrow 1/2^-_1$    &  4.2(7)    &  2.4    \\

\\
$^{197}$Hg   & $5/2^-_1 \rightarrow 1/2^-_1$      &  8.74(25)    &  2.9  \\
             & $(3/2)^-_1 \rightarrow 5/2^-_1$    &  NA          &  4.1  \\
             & $(3/2)^-_1 \rightarrow 1/2^-_1$    &  NA          &  1.7   \\
             
\\
$^{199}$Hg  & $5/2^-_1 \rightarrow 1/2^-_1$     &  17.6(3)     &  11.5     \\
            & $3/2^-_1 \rightarrow 5/2^-_1$     &  12.5(25)    &  4.2   \\
            & $3/2^-_1 \rightarrow 1/2^-_1$     &  16.1(11)    &  12.4   \\
            
\hline
\hline
 \label{t_be2odd}
\end{tabular}
\end{table*}

\begin{table*}
\centering
\caption{Comparison of calculated and experimental \cite{NNDC, mu_data_table, iaea} quadrupole ($Q$)  and magnetic moments ($\mu$) of odd-$A$ Hg isotopes in the units of $e$b and $\mu_N$, respectively. }
\begin{tabular}{ lcccccccccc } 
 \hline
 \hline
& \hspace{2cm}   &  \multicolumn{2}{c}{$Q\ (e$b$)$} &  & \multicolumn{2}{c}{$\mu\ (\mu_{N})$ }& & \multicolumn{1}{c}{$\mu^{\prime}\ (\mu_{N})$ }  \\
 \cline{3-4}
 \cline{6-7}
 \cline{9-9}
Nucleus &$J^{\pi}$&  Expt. & SM & \hspace{0.5cm} & Expt. & SM &  & Lit. \cite{mu_oddHg} \\
 \hline

\\
 $^{193}$Hg     & 1/2$_1^-$     &  $-$       &  $-$       & &  NA          & 0.30  &   & $-$  \\
 
 	          & 5/2$_1^-$     &  NA        & -0.76      & &  NA          & 0.48  &      & $-$  \\
 	          
 	          & 3/2$_1^-$     &  -0.7(3)   & -0.43      & &  -0.6251(8)  & -0.53 &     & -0.44  \\
 	          
 	          & 13/2$^+$      &  0.92(2)   & 1.06       & &  -1.0543(12) & -0.77 &      & -1.25  \\
\\
$^{195}$Hg      & 1/2$_1^-$    &  $-$       &  $-$       & &  0.5393(6)   & 0.30 &       & 0.45 \\

 	          & 5/2$_1^-$    &  NA        & -0.70      & &  NA          & 0.52 &       & $-$ \\
 	          
 	          & 3/2$_1^-$    &  NA        & -0.43      & &  NA          & -0.47 &      & $-$ \\
 	          
 	          & 13/2$^+$     &  1.08(2)   & 1.20       & &  -1.0405(12) & -0.75 &      & -1.25 \\
 \\
$^{197}$Hg      & 1/2$_1^-$    &  $-$        &  $-$       & &  0.5253(6)   & 0.30 &      & 0.43 \\

 	          & 5/2$_1^-$    &  -0.081(6)  &  -0.46      & &  0.855(15)   & 0.60 &       & $-$ \\
 	          
 	          & 3/2$_1^-$    &  NA         &  -0.30      & &  NA          & -0.52 &     & $-$  \\
 	          
 	          & 13/2$^+$     &  1.25(3)    &  1.35      & &  -1.0236(12) & -0.72 &     & -1.44 \\
\\
$^{199}$Hg      & 1/2$_1^-$    &  $-$      &  $-$       & &  0.5039(6)   & 0.29  &       & 0.44  \\

 	          & 5/2$_1^-$    &  0.95(7)  &  0.45      & &  0.88(3)     & 0.76 &   & $-$ \\
 	          
 	          & 3/2$_1^-$    &  0.62(15) &  0.41      & &  -0.56(9)    & -0.42 &       & $-$ \\
 	          
 	        
 	          & 13/2$^+$     &  1.2(3)   &  1.28      & &  -1.0107(12) & -0.72 &       & -1.44 \\
 	          
 \\

 \hline
 \label{t_qmOdd}
\end{tabular}
\end{table*}


The shell-model results for electric quadrupole ($Q$) and magnetic moments ($\mu$) of the first three low-lying states ($1/2_1^-$, $3/2_1^-$, and $5/2_1^-$) are shown in Table \ref{t_qmOdd}. Both the shell model and the measured quadrupole moments of the states $3/2_1^-$ and $5/2_1^-$ have the same order of magnitude and signs. We obtained a large difference between the calculated and experimental data only in the $Q$ value of $5/2_1^-$ state in $^{197}$Hg. In other cases, the magnitudes of the quadrupole moments are reduced compared to the experimental data in the shell-model results. We have also provided shell-model predictions for quadrupole moments of $3/2_1^-$ and $5/2_1^-$ states in odd -$A$ Hg isotopes where experimental data are not available yet. The quadrupole moments of the isomeric state $13/2_1^+$ are well reproduced by the shell model.
The shell-model $Q$ values can serve as the basis for comparing the quadrupole moments predicted from different theoretical models and upcoming experimental data.


The magnetic moments of the $1/2_1^-$, $3/2_1^-$, and $5/2_1^-$ states given by the shell model are in reasonable agreement with the experimental data. The magnetic moments of odd-mass Hg isotopes are recently evaluated in Ref. \cite{mu_oddHg} using the mean-field approach. In Ref. \cite{mu_oddHg}, the Hartree-Fock-Bogoliubov (HFB) method is implemented to obtain ground-state configuration compatible with the observed spin, parity, and the measured magnetic moments, by blocking the unpaired nucleon in a self-consistent way. Then the magnetic moments are calculated for which experimental data are available. The $\mu$ values calculated in the literature \cite{mu_oddHg} (named as $\mu^{\prime}$ hereafter) are compared with the theoretical predictions from the shell model. The magnetic moments ( or $\mu^{\prime}$ ) in Table \ref{t_qmOdd} are extracted from Fig. 3 of Ref. \cite{mu_oddHg} without considering the theoretical uncertainties. 
We observed that the shell-model $\mu$ values of the isomeric state $13/2_1^+$ are overestimated while those reported in Ref. \cite{mu_oddHg} are underestimated. However, the deviations from the measured value are approximately the same in both cases. The shell-model $\mu$ of the $3/2_1^-$ state in $^{193}$Hg is more closer to the experimental value than $\mu^{\prime}$. The $\mu$ values of the $ 1/2_1^- $ state remain almost constant in both models in the Hg chain. The magnetic moments of the $5/2_1^-$ state are not discussed in \cite{mu_oddHg}. We have shown the shell-model predictions for it along with the available experimental data. 

The positive magnetic moments in the $ 1/2_1^- $ and $5/2_1^-$ states indicate the dominance of neutron configurations influenced by the $j_{<}$ orbital while the negative values in the $ 3/2_1^- $ and $13/2_1^+$ states imply the dominance of neutron configurations of the $j_{>}$ orbital. The deviation of the measured magnetic moments from the single particle value (or Schmidt values), evaluated with the effective $g$-factors used here, is not so large. For $ 1/2_1^- $ and $5/2_1^-$, the deviation is $\sim$ 0.2 and 0.15 units, respectively, while in the case of $ 3/2_1^- $ state, they differ by $\sim$ 0.4 units. The positive shift from the Schmidt value can be partly attributed to proton contributions since they are associated with 
 positive $g$-factors.
On the other hand, the measured $\mu$ of the $13/2_1^+$ level closely aligns with the Schmidt value (-1.0025 $\mu_{N}$) and differs by $\sim$ -0.3 units from the calculated values. This may suggest the need for the refinement of the effective $g$- factors of the nucleons, particularly for the $g_l$ of neutron ($g_l^{n}$). Since the magnetic moment of $13/2_1^+$ is very sensitive to $g_l^{n}$, inclusion of a small $g_l^{n}$ about -0.05 improves magnetic moments of the $13/2_1^+$ levels to $\sim$ -1.0 $\mu_{N}$. This adjustment also enhances the magnitudes of magnetic moments of the $10_1^+$ and $12_1^+$ states corresponding to  Table \ref{t_qmEven} results,
bringing them closer to the experimental data.

\section{Conclusions}
\label{sec5}

We have performed a systematic study of $^{193-200}$Hg isotopes in the shell-model framework. The shell-model calculations were carried out in the $50 \leq Z \leq 82$ and  $82 \leq N \leq 126$ model space with monopole-based truncation using the KHHE interaction. In the even-even Hg isotopes, the calculated energy levels and the energy systematic of the positive and negative parity yrast states are in fair agreement with the corresponding experimental data and indicate $\nu 0i_{13/2}$ sub-shell closure in $^{200}$Hg. The calculated $E2$ strengths among various transitions are compared with the recently measured data. The change in dominant configurations of the positive parity yrast states with spin is analyzed based on shell-model wave functions, and their impact on $E2$ matrix elements is discussed. The quadrupole and magnetic moments of the low-lying and isomeric states are evaluated by the shell model. A similar discussion is followed for energy spectra and electromagnetic moments in the case of odd-mass Hg isotopes.
The experimental low-lying spectra of odd-mass Hg isotopes are well reproduced by the shell-model results, which demonstrate that the low-lying states are interpreted as zero, one, two $\nu0i_{13/2}$-quasiparticle states. Exceptionally, the $1/2^-$ ground-state spin cannot be reproduced by the shell-model calculation with the monopole-based truncation, the QVSM calculation demonstrated that the calculation without any truncation would give the correct ordering of the low-lying three states.   
In some cases, the shell-model results for Hg isotopes are also compared with the available results of other theoretical models, such as IBM and HFB methods. On average,  the shell-model results reasonably agree with the experimental data. We have provided the shell-model predictions of these observables for different states where experimental data are unavailable. These shell-model results can serve as a useful guide for upcoming experiments and might even inspire future experiments. The energy surfaces of Hg isotopes indicate oblate deformation, transitioning gradually towards a spherical shape at $^{200}$Hg.

\section*{Acknowledgments}
S.S. would like to thank UGC (University Grant Commission), India for financial support for his Ph.D. thesis work. We acknowledges a research grant from SERB (India), CRG/2022/005167. We would like to thank the National Supercomputing Mission (NSM) for providing computing resources of ‘PARAM Ganga’ at the Indian Institute of Technology Roorkee, implemented by C-DAC and supported by the Ministry of Electronics and Information Technology (MeitY) and Department of Science and Technology (DST), Government of India.   P.C.S. acknowledges the hospitality extended to him during his stay at the Center for Computational Sciences, University of Tsukuba, Japan. N.S. and Y. U. acknowledge the support of ``Program for promoting researches on the supercomputer Fugaku'', MEXT, Japan (JPMXP1020230411) and the MCRP program of the Center for Computational Sciences, University of Tsukuba (wo22i002, NUCLSM).

\end{document}